\documentclass[a4paper,11pt]{article}
\pdfoutput=1

\usepackage{jheppub}
\usepackage{graphicx}
\usepackage{amsmath}
\usepackage{bm}
\usepackage{cancel}
\usepackage{booktabs}
\usepackage[small]{subfigure}
\usepackage{url}

\newcommand{\eq}[1]{eq.~\eqref{eq:#1}}
\newcommand{\eqs}[2]{eqs.~\eqref{eq:#1} and \eqref{eq:#2}}
\renewcommand{\sec}[1]{section~\ref{sec:#1}}

\newcommand{\subsec}[1]{section~\ref{subsec:#1}}

\newcommand{\subsecs}[2]{sections~\ref{subsec:#1} and \ref{subsec:#2}}

\newcommand{\fig}[1]{figure~\ref{fig:#1}}

\newcommand{\mycite}[1]{ref.~\cite{#1}}
\newcommand{\mycites}[1]{refs.~\cite{#1}}

\newcommand{\ord}[1]{\mathcal{O}(#1)}
\newcommand{\Ord}{\mathcal{O}}

\newcommand{\mylog}[1]{\ln\left( #1 \right)}

\newcommand{\MSb}{\overline{\rm MS}}

\newcommand{\muf}{\mu_F}
\newcommand{\mur}{\mu_R}

\newcommand{\mub}{\mu_b}
\newcommand{\as}{\alpha_s}

\newcommand{\mh}{m_H}
\newcommand{\mb}{m_b}

\newcommand{\GeV}{\,\mathrm{GeV}}

\newcommand{\nn}{\nonumber}

\newcommand{\nfour}{{{[4]}}}
\newcommand{\nfive}{{{[5]}}}

\newcommand{\pa}{\mathrm{partial}}

\newcommand{\NLO }{NLO{\footnotesize[$Y_b^2$]}}
\newcommand{\NLOt}{NLO{\footnotesize[$Y_b^2\!+\!Y_bY_t$]}}

\allowdisplaybreaks[2]

\title{\boldmath Matched predictions for the $b\bar{b}H$ cross section at the 13 TeV LHC}

\author[a]{Marco Bonvini,}
\author[b]{Andrew S.~Papanastasiou,}
\author[c]{and Frank J.~Tackmann}
\affiliation[a]{Rudolf Peierls Center for Theoretical Physics,\\1 Keble Road, University of Oxford, OX1 3NP Oxford, UK}
\affiliation[b]{Cavendish Laboratory, University of Cambridge,\\J.J. Thomson Avenue, CB3 0HE, Cambridge, UK}
\affiliation[c]{Theory Group, Deutsches Elektronen-Synchrotron (DESY),\\Notkestra\ss e 85, D-22607 Hamburg, Germany}

\emailAdd{marco.bonvini@physics.ox.ac.uk}
\emailAdd{andrewp@hep.phy.cam.ac.uk}
\emailAdd{frank.tackmann@desy.de}

\abstract{
We present up-to-date matched predictions for the $b\bar{b}H$ inclusive cross section at the LHC at $\sqrt{s}=13$~TeV.
Using a previously developed method,
our predictions consistently combine the complete NLO contributions that are present in the 4-flavor scheme calculation,
including finite $b$-quark mass effects as well as top-loop induced $Y_b Y_t$ interference contributions,
with the resummation of collinear logarithms of $m_b/m_H$ as present in the 5-flavor scheme calculation up to NNLO.
We provide a detailed estimate of the perturbative uncertainties of the matched result by examining its dependence on the
factorization and renormalization scales, the scale of the Yukawa coupling, and also the low $b$-quark matching scale in the PDFs.
We motivate the use of a central renormalization scale of $m_H/2$, which is halfway between the values typically chosen in the 4-flavor and 5-flavor scheme calculations.
We evaluate the parametric uncertainties due to the PDFs and the $b$-quark mass, and in particular
discuss how to systematically disentangle the parametric $m_b$ dependence and the unphysical $b$-quark matching scale dependence.
Our best prediction for the $b\bar bH$ production cross section in the Standard Model at $13$~TeV and for $m_H = 125\GeV$
is $\sigma(b\bar b H) = 0.52 \,{\rm pb}\, \bigl[1 \pm 9.6\% {\rm (perturbative)}\,{}^{+2.9\%}_{-3.6\%} {\rm (parametric)} \bigr]$.
We also provide predictions for a range of Higgs masses $m_H\in [50, 750]\GeV$.
Our method to compute the matched prediction and to evaluate its uncertainty
can be readily applied to other heavy-quark-initiated processes at the LHC.
}


\preprint{
\begin{flushright}
OUTP-16-08P\\
CAVENDISH-HEP-16-06 \\
DESY 16-076\\
May 5, 2016
\end{flushright}
}

\begin{document}

\maketitle

\section{Introduction}
\label{sec:intro}

Heavy-quark initiated processes and accurate theoretical predictions for them are important
for precision tests of the Standard Model (SM), searches for
New Physics, and are also essential for the precise determination of parton distribution functions (PDFs).

Predictions for processes involving initial-state $b$-quarks at hadron colliders are typically made
using factorization theorems that are valid for two different parametric
hierarchies between the $b$-quark mass, $m_b$, and the hard-interaction scale of the process, $Q$.
In the 4-flavor scheme (4FS), one formally considers $m_b \sim Q$, while in the 5-flavor scheme (5FS)
one formally considers $m_b \ll Q$. The predictions obtained in either scheme have different features.
The 4FS predictions include power corrections in $m_b/Q$ as well as the exact massive quark final-state
phase space, while logarithms of $\mylog{m_b/Q}$ are included at a given fixed order. The 5FS predictions
do not include power corrections in $m_b/Q$ but resum the potentially
large logarithms $\mylog{m_b/Q}$ to all orders via DGLAP into a $b$-quark PDF.
The construction of matched predictions that include the merits of both schemes in DIS
(sometimes called variable flavor number schemes) has been the subject of many years of work
\cite{Aivazis:1993kh, Aivazis:1993pi, Thorne:1997ga, Kretzer:1998ju, Collins:1998rz, Cacciari:1998it, 
Kramer:2000hn,Tung:2001mv, Thorne:2006qt, Forte:2010ta, Guzzi:2011ew, Hoang:2015iva, Ball:2015dpa} and 
recently progress in this direction for hadron-hadron colliders has also been 
made~\cite{Han:2014nja, Forte:2015hba, Bonvini:2015pxa}.

The focus of this paper is on $b\bar{b}H$ production, for which predictions for the 
inclusive cross section are available up to NNLO in the 5FS~\cite{Dicus:1998hs,Balazs:1998sb,Harlander:2003ai,Buehler:2012cu,Harlander:2012pb}
and NLO in the 4FS~\cite{Dittmaier:2003ej,Dawson:2003kb,Wiesemann:2014ioa}.
In the past, the results in the two schemes have been averaged using
the pragmatic Santander-Matching prescription~\cite{Harlander:2011aa}.
Given that this prescription amounts to a simple weighted average of the two results, it is
clear that it does not constitute a satisfactory and theoretically-consistent matching.
In particular, as already observed in \mycite{Bonvini:2015pxa},
the contributions present in one and not the other scheme should get added
in their combination rather than averaged, which in this case leads to a noticeably
higher cross section of the properly matched result compared to the Santander average.

In \mycite{Bonvini:2015pxa}, we used a simple effective field theory setup to systematically
derive matched predictions for the $b\bar b H$ cross section, which combines the ingredients of
both 4FS and 5FS schemes. The result contains the full 4FS result, including the exact
dependence on the $b$-quark mass, and improves it with a resummation of collinear logarithms of $\mb/\mh$,
as are present in the 5FS.
An important aspect of our construction is the perturbative counting of the effective 
$b$-quark PDF, which is counted as an $\Ord(\as)$ object for phenomenologically relevant
hard (factorization) scales below $\sim 1$ TeV. 
This counting rearranges the perturbative expansion of the cross section, such
that in the massless limit the result corresponds to a reorganized 5FS result.
An important advantage is that in this way the logarithms present in the resummed (5FS) result
order-by-order exactly match those present in the fixed-order (4FS) contributions.
This feature greatly facilitates the combination of both types of contributions.

In this paper, using the approach of \mycite{Bonvini:2015pxa},
we present state-of-the-art predictions for the $b\bar{b}H$ cross section for the LHC at 13~TeV,
including a comprehensive study of its perturbative and parametric uncertainties.
Our predictions include the effects of top-quark loop-induced interferences,
proportional to $Y_bY_t$, which are known to be important in the Standard Model
\cite{Dittmaier:2003ej,Dawson:2003kb,Wiesemann:2014ioa}. (These were not yet included in
\mycite{Bonvini:2015pxa}.)
We also investigate in detail the effects of pure 2-loop terms that are present in the 5FS NNLO calculation
but are formally of higher order in our perturbative counting.

The two-step matching used in \mycite{Bonvini:2015pxa} makes it explicit that there are
two relevant scales in the problem: the hard scale $\mu_H \sim \mh$
and the $b$-quark scale $\mu_b \sim m_b$ (also referred to as the $b$-quark threshold scale).
Since a perturbative expansion is performed at both of these scales, a reliable theory uncertainty should take into
account the perturbative uncertainties related to both.
The uncertainty due to $\mu_b$ has been neglected in the past in essentially all 5FS and matched cross section predictions, since all
standard 5FS PDF sets make the fixed choice $\mu_b=m_b$.
However, the $\mu_b$ uncertainty should be regarded as an additional resummation uncertainty, and
in \mycite{Bonvini:2015pxa} it was shown how to systematically estimate it and that it can indeed
have a nonnegligible effect.
We discuss and motivate an appropriate choice of the central scales and provide a full
breakdown of theoretical uncertainties due to $\mu_F$, $\mu_R$, $\mu_b$ variations.

We show that this more general setup also disentangles the dependence on $\mu_b$ and $m_b$
and thus allows one to correctly evaluate the uncertainty in the predictions due to the parametric uncertainty in the value of $m_b$.
This discussion is relevant for any heavy-quark initiated process
calculated in either the fixed 5FS or a matched approach.

The structure of the paper is as follows: in \sec{nlonll}, we recall the main features in
the construction of the matched NLO+NLL result and discuss the extensions to include
the $Y_bY_t$ interference terms and higher-order 2-loop terms.
In \sec{uncertainty}, we discuss in detail the perturbative and parametric uncertainties.
In \sec{lhc13}, we present our numerical results for the inclusive cross section for a range of
Higgs masses $m_H \in [50,750]\GeV$, including the full breakdown of all uncertainties.
We conclude and give a short outlook in \sec{conclusions}.

\section{Matched cross section}
\label{sec:nlonll}

In this section we discuss the theoretical ingredients for our predictions of the
$b\bar bH$ cross section.
In \subsecs{FOresum}{matching}, we summarize the main steps and results of \mycite{Bonvini:2015pxa}
in the construction of the fixed-order + resummation matched result, which is valid for any parametric
hierarchy between $m_b$ and $m_H$. For the detailed derivation we refer to \mycite{Bonvini:2015pxa}.
In \subsec{nlonnllp}, we discuss the inclusion of the formally higher-order two-loop terms in the
resummed results. We will denote the corresponding result as NLO+NNLL$_\pa$.
In \subsec{ybyt}, we discuss the inclusion of the $Y_b Y_t$ interference terms.

\subsection{Fixed order and resummed results}
\label{subsec:FOresum}

The fixed-order cross section that enters the matched result, and which is recovered when the resummation is turned off,
is obtained in a factorization scheme that is formally valid for the parametric power counting $m_b \sim m_H$.
Hence, it is equivalent to a 4FS result, where bottom quarks do not appear in the initial-state and can only be produced via gluon splittings into $b$-quark pairs. It includes the exact dependence on the $b$-quark mass (i.e., it includes power corrections to all orders in $m_b^2/m_H^2$) both in the partonic cross sections and in the phase space.
Logarithmic terms $\sim\ln(m_b^2/m_H^2)$ arising from collinear gluon splittings
are included at fixed order in $\as$.
The fixed-order result is given in terms of coefficient functions
$D_{ij}(m_H,m_b,\muf)$, which depend explicitly on $m_b$, and 4-flavor PDFs, 
\begin{align} \label{eq:bbH4FSexplicit}
  \sigma^{\rm FO} &= \sum_{i,j=g,q,\bar q} D_{ij}(m_H,m_b,\muf) \,
                     f^{[4]}_i(\muf)\, f^{[4]}_j(\muf)
\nn\\
                  &= \sum_{i,j,k,l=g,q,\bar q}  D_{ij}(m_H,m_b,\muf)
                    \Bigl[U^{[4]}_{ik}(\muf,\mu_0) f^{[4]}_k(\mu_0)\Bigl]
                    \Bigr[U^{[4]}_{jl}(\muf,\mu_0) f^{[4]}_l(\mu_0)\Bigr]
\,.\end{align}
For notational simplicity, we keep all Mellin convolutions between PDFs, evolution factors, and coefficient functions implicit and simply write products throughout the paper.
The sum over partons only includes gluons and light (anti)quarks ($q=d,u,s,c$),
$\muf$ is the hard factorization scale of the process. In the second line
the 4-flavor PDFs $f^{[4]}_j(\muf)$ are written in terms of the fitted PDFs at the low scale $\mu_0$%
\footnote{In this work we consider the charm quark to be a fitted light-quark PDF. In most PDF fits, the charm PDF, 
as with the bottom PDF, is generated perturbatively, however this is not of relevance for the 
present discussion.} and DGLAP evolution factors $U^{[4]}_{ij}$ with $n_f = 4$ active quark flavors.
The fixed-order result of \eq{bbH4FSexplicit} is equivalent to a 4FS result and the coefficients $D_{ij}$ for $b\bar{b}H$
are known to NLO.

The pure resummed result is based on the factorization theorem derived in the limit $m_b \ll m_H$, i.e., in a
power expansion in $m_b/m_H$, where the leading power term leads to the usual 5FS result, while subleading power
terms correspond to subleading twist terms and are usually not considered.
In this case, $b$-quarks are treated as massless at the hard matching scale $\muf\sim m_H$ and appear in the
initial state of the corresponding partonic matching calculation.
The collinear logarithms of $m_b/m_H$ are resummed by DGLAP evolution from the hard scale $\muf\sim m_H$ to the
$b$-quark matching scale $\mub \sim m_b$. At $\mub$, the $b$-quark is integrated out of the theory.
These steps result in an effective perturbative $b$-quark PDF.
The resummed cross section is given by the convolution of coefficient functions $C_{ij}(m_H,\muf)$, which
contain no $m_b$ dependence, with 5-flavor PDFs,
\begin{align}
  \label{eq:bbH5FSexplicit}
  \sigma^{\rm resum} &= \sum_{i,j=g,q,\bar q,b,\bar b} C_{ij}(m_H,\muf)\,
                     f^{[5]}_i(m_b,\muf) \, f^{[5]}_j(m_b,\muf) 
\\\nonumber
  &= \sum_{i,j=g,q,\bar q,b,\bar b} C_{ij}(m_H,\muf)
  \Biggl[\sum_{\substack{k=g,q,\bar q,b,\bar b\\l,p=g,q,\bar q}} U^{[5]}_{ik}(\muf,\mub) \mathcal{M}_{kl}(m_b,\mub)
    U^{[4]}_{lp}(\mub,\mu_0) f^{[4]}_p(\mu_0) \Biggr]
\nonumber\\\nonumber & \qquad\qquad\qquad\qquad\qquad\times
\Biggl[\sum_{\substack{k=g,q,\bar q,b,\bar b\\l,p=g,q,\bar q}} U^{[5]}_{jk}(\muf,\mub) \mathcal{M}_{kl}(m_b,\mub)
    U^{[4]}_{lp}(\mub,\mu_0) f^{[4]}_p(\mu_0)\Biggr]
\,.\end{align}
In the second step, the 5-flavor PDFs at $\mu=\muf$ are written out explicitly
in terms of the 5-flavor evolution from $\muf$ to $\mub$, the matching at $\mub$ yielding matching coefficients $\mathcal{M}_{kl}(m_b,\mub)$ that explicitly depend on $m_b$, followed by the 4-flavor evolution from $\mub$ to $\mu_0$ and the fitted 4-flavor PDFs at scale $\mu_0$.
All mainstream 5FS PDF sets construct their $b$-quark PDFs in this way, however, with the fixed choice of $\mub = m_b$.
With this choice, the $\Ord(\as)$ matching coefficients $\mathcal{M}_{ij}$ in the $m_b$ pole-scheme are exactly zero,
which somewhat simplifies the implementation.
However, identifying $\mub = m_b$ confuses the parametric physical dependence on $m_b$ and the unphysical dependence on the matching scale
$\mub$, which controls the resummation of logarithms and should cancel to the order one is working.
We will discuss how we rectify this situation in \sec{uncertainty}.

\subsection{Matching fixed order and resummation: NLO+NLL}
\label{subsec:matching}

\begin{figure}[t]
\centering
\includegraphics[width=0.8\textwidth]{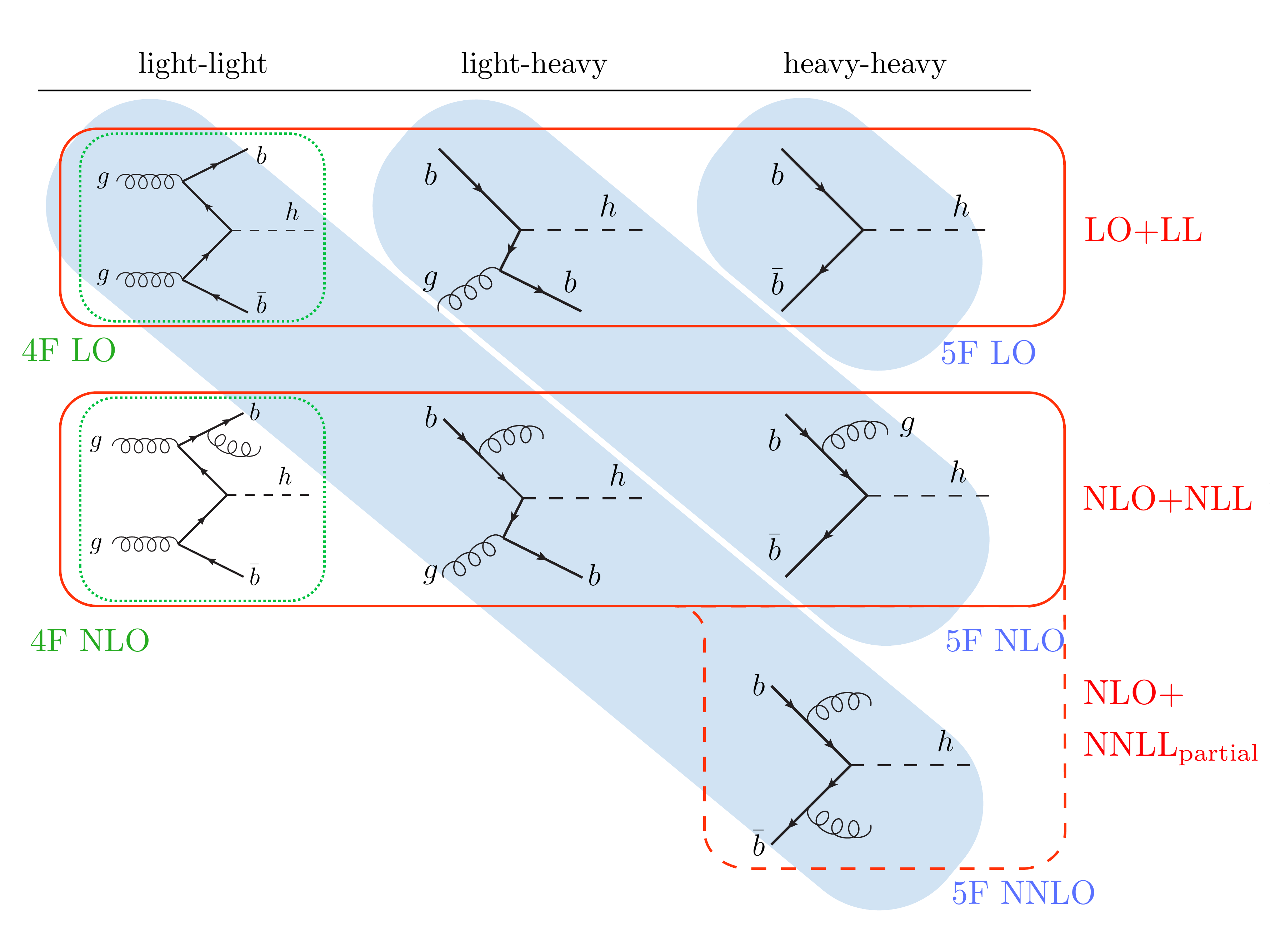}
\caption{Pictorial representation with sample diagrams appearing in the computation of the $b\bar bH$ cross section,
  grouped according to the different perturbative countings adopted in the 4FS (green boxes), 5FS (blue areas)
  and our matched resummed result (red boxes).}
\label{fig:bbh-diagrams}
\end{figure}

In all practical applications we are aware of, the \emph{evolved} PDFs are always treated as external
$\Ord(1)$ objects and the perturbative expansion of the cross section is based solely on the perturbative expansion of
the coefficient functions $D_{ij}$ and $C_{ij}$ in \eqs{bbH4FSexplicit}{bbH5FSexplicit}.
For the fixed-order and resummed results this leads to the usual  4FS and 5FS predictions.
The corresponding contributions are schematically depicted in \fig{bbh-diagrams} by the first column for the 4FS
(green dotted boxes) and by the three diagonals (blue shading) for the 5FS.

As noted above, the $b$-quark PDF is itself a perturbative object with the expansion
\begin{align} \label{eq:5FPDF_exp_resum}
f^{[5]}_b(m_b, \muf)
&=
\Bigl[U_{bg}^\nfive(\muf,\mub)
+ \frac{\as(\mub)}{4\pi}\, U_{b b}^\nfive(\muf,\mub) \mathcal{M}_{bg}^{(1)}(\mb, \mub) \Bigr] f_g^\nfour(\mub)
+\, \dotsb
\nn\\*
&\sim \qquad \ord{\as} \quad + \quad\qquad \qquad\qquad \ord{\alpha_s} \qquad\qquad\qquad\qquad + \ord{\as^2}
\,.\end{align}
In principle, each of its terms should be included in the perturbative counting and be regarded as part of the perturbative expansion of the cross section. Counting it as an $\Ord(1)$ object would be justified in the limit where the off-diagonal evolution factor
$U^{[5]}_{bg}(\muf,\mub) \sim 1$.
However, $U^{[5]}_{bg}(\muf, \mub)$ is suppressed by an overall $\as \ln(\muf/\mub)$ relative to the diagonal evolution
factors and vanishes in the limit $\mub\to \muf$, and therefore this only holds for scales $\muf \ggg \mub$. 
Numerically, for $\mub\sim m_b$ this is only attained for scales $\muf \gtrsim 1$~TeV. 
Hence, for the scales of interest here it is more appropriate to count $U^{[5]}_{bg}(\muf, \mub)$
as $\Ord(\as)$, as indicated in \eq{5FPDF_exp_resum}, in which case the whole $f_b^{[5]}$ becomes an $\Ord(\as)$ object.
The perturbative expansion in $\as$ of the resummed cross section in \eq{bbH5FSexplicit} is then performed
by expanding the product of coefficient functions $C_{ij}$ \emph{together} with the
terms making up the $b$-quark PDF including the $b$-quark matching coefficients $\mathcal{M}_{ij}$ and $U_{bg}^{[5]}\sim\as$.
For a more detailed discussion we refer to \mycite{Bonvini:2015pxa}.

The 4FS and 5FS results can significantly differ from each other and in particular display different
patterns in their factorization-scale dependence due to the different logarithmic terms present
at each order in the two schemes. Hence, a consistent matching appears to be nontrivial.
The above treatment of the resummed result has the important added advantage that
it reorganizes the resummed series into a form that is consistent with the logarithms present in the fixed-order result.
The key feature is that order by order in $\as$ the limit $\mub \to \muf$ in the resummed cross section 
now exactly reproduces \emph{all} the logarithmic terms (and nothing more) that are present in the
$m_b\to 0$ limit of the fixed-order cross section. 
In other words, the reexpansion of the resummed result to fixed order is simply given by setting $\mub = \muf$.
This in turn means that for $\mub < \muf$ the evolution factors $U^{[5]}_{ij}$ in this expansion precisely resum 
the singular logarithms present in the fixed-order result.
Hence, all that is missing in the resummed result compared to the fixed-order result are purely nonsingular terms
proportional to $m_b^2/\mh^2$, i.e.\ terms that vanish in the
limit $m_b \to 0$ given by
\begin{align}
\sigma^{\rm nons}
= \sigma^{\rm FO} - \sigma^{\rm resum}\big|_{\mub=\muf}
\,.\end{align}
The complete matched cross section is then simply given by adding
the nonsingular fixed-order terms to the resummed result,
\begin{align} \label{eq:matching}
\sigma^{\rm FO+resum} = \sigma^{\rm resum} + \sigma^{\rm nons}
= \sigma^{\rm resum} + \bigl( \sigma^{\rm FO} - \sigma^{\rm resum}\big|_{\mub=\muf} \bigr)
\,.\end{align} 
By construction, it satisfies $\sigma^{\rm FO+resum} \to \sigma^{\rm FO}$
in the limit $\mub\to \muf$ where the resummation is turned off, as required for a consistently matched prediction.
On the other hand, it reduces to $\sigma^{\rm resum}$ in the limit
$m_H\gg m_b$.  We emphasize again that this crucially relies on the
fact that the nonsingular terms vanish for $m_b\to 0$, which in turn
relies on adopting the perturbative counting for the resummed result
described above.%
\footnote{Other approaches combining resummed and fixed-order
  expressions proceed similar to \eq{matching}.  However, if a
  perturbative counting different from the one we use is adopted, the
  singular contributions that are common to both cannot be obtained by
  simply setting $\mub=\muf$. In this case (see for example
  \mycites{Cacciari:1998it, Forte:2010ta}) the singular terms can be
  computed by explicitly expanding the resummed result in powers of
  $\as$, but only those terms which are present in the fixed-order
  result and would otherwise be double counted are subtracted. The
  so-matched result will however not reproduce the fixed-order result
  in the limit $\mub\to\muf$, since the resummed result and the
  singular subtractions will not cancel each other in this limit.}
The corresponding terms included in the matched result at each order
are depicted by the rows (red boxes) in \fig{bbh-diagrams}.

As discussed in \mycite{Bonvini:2015pxa}, for the practical implementation, the nonsingular contributions can
be conveniently absorbed into modified gluon and light-quark coefficient functions, $\bar{C}_{ij}(m_H,m_b,\muf)$, 
which now carry an explicit dependence on $m_b$, convolved with effective 5F PDFs.%
\footnote{Moving the nonsingular corrections underneath the 5F resummation corresponds to including some 
resummation effects for power-suppressed terms, which is beyond the formal accuracy in either the 4FS or 5FS.}
The final matched result is then written as
\begin{align}\label{eq:bbHMatchedexplicit}
\sigma^{\rm FO+resum}
&= \sum_{i,j=b,\bar b} C_{ij}(m_H,\muf) \, f^{[5]}_i(m_b,\muf)\, f^{[5]}_j(m_b,\muf)
\nonumber\\ & \quad
+ \sum_{\substack{i=b,\bar b\\ j=g,q,\bar q}}
\Bigl[C_{ij}(m_H,\muf) \, f^{[5]}_i(m_b,\muf)\, f^{[5]}_j(m_b,\muf)
+ (i \leftrightarrow j)  \Bigr]
\nonumber\\ & \quad
+ \sum_{i,j=g,q,\bar q} \bar C_{ij}(m_H,m_b,\muf) \, f^{[5]}_i(m_b,\muf) \, f^{[5]}_j(m_b,\muf)
\,,\end{align}
where $f^{[5]}_{i,b}$ are perturbative objects, and an expansion of $C_{ij}$ and $\bar{C}_{ij}$ against $f^{[5]}_{i,b}$ 
as discussed above is implicit.

The strict expansion of the coefficient functions against the individual terms making up
the $b$-quark PDF is quite inconvenient for the practical implementation, as it requires performing the entire $n_f = 5$ DGLAP evolution above $\mu_b$ by hand. However, as long as we are only interested in the phenomenologically
relevant region $\mu_b\sim m_b \ll \muf \sim m_H$, we can also keep formally higher-order cross terms in order to
simplify the practical implementation. Doing so allows us to use common preevolved 5FS PDFs, under the condition that
we count $f^{[5]}_b \sim \Ord(\as)$ while the light quark and gluon PDFs are counted as $\sim \ord{1}$. In addition, we have to use
PDFs of sufficiently high order such that they include all matching corrections required by our perturbative counting.
Specifically, at (N)LO+(N)LL this requires the use of at least (N)NLO PDFs.
It was explicitly checked in \mycite{Bonvini:2015pxa} that for values of $m_b/m_H \lesssim 0.1$ this implementation
gives practically the same numerical results as the strict expansion. On the other hand, the strict expansion is required if one wishes to explicitly study the limit $\mu_b\to\muf$ and obtain a smooth transition of the matched result into the fixed-order result.

In summary, with this simplification, expanding the matched cross section in powers of $\as=\as(\muf)$, the following
perturbative expansion is obtained
\begin{align} \label{eq:bbhCountingRes}
 \text{LO+LL} && \sigma &= \quad \as^2 \bar C^{(2)}_{ij}\, f_{i}^{[5]} f_{j}^{[5]}
    + \as 4 C^{(1)}_{bg}\, f_b^{[5]} f_g^{[5]}
    + \phantom{\as}2C^{(0)}_{b\bar b}\, f_b^{[5]} f_b^{[5]}
  \nonumber\\
 \text{NLO+NLL} &&  & \quad + \as^3 \bar C^{(3)}_{ij}\, f_{i}^{[5]} f_{j}^{[5]}
    +\as^2 4 C^{(2)}_{bk}\, f_b^{[5]} f_k^{[5]}
    +\as 2C^{(1)}_{b\bar b}\, f_b^{[5]} f_b^{[5]}
  \nonumber\\
 \text{NNLO+NNLL} &&  & \quad + \as^4 \bar C^{(4)}_{ij}\, f_{i}^{[5]} f_{j}^{[5]}
    +\as^3 4 C^{(3)}_{bk}\, f_b^{[5]} f_k^{[5]}
    +\as^2 (2C^{(2)}_{b\bar b}+2C^{(2)}_{bb}) \, f_b^{[5]} f_b^{[5]}
  \nonumber\\
 && & \quad + \ord{\as^5}
\,.\end{align}
The factors of two and four account for the exchange of partons among the two protons and (to a first approximation) the equality $f_b^{[5]}=f_{\bar b}^{[5]}$.
A sum over \emph{light} quarks and gluons is implicitly assumed for repeated   indices $i,j,k$.
The superscripts on the coefficient functions indicate the order in $\as$ to which these are computed.
The first two orders in \eq{bbhCountingRes} are illustrated by the red boxes in \fig{bbh-diagrams}.
As seen there, our perturbative counting implies that we include $b\bar{b}$, $bg$ and $gg$ initiated contribution consistently at the same order.
 
Finally, we note that the construction of the coefficient functions $\bar{C}_{ij}$ is
formally the same as the corresponding construction in the FONLL approach~\cite{Forte:2015hba}
(and in a hypothetical S-ACOT construction). There are, however, two main differences between these approaches. 
First, as explained above, we use the fact that the effective $b$-quark PDF counts as an $\Ord(\as)$ perturbative object
to construct the perturbative expansion of our matched result.
As a result, it contains the complete fixed-order result at each perturbative order, and (with the strict expansion)
smoothly merges into it.
Secondly, as discussed further in \sec{uncertainty}, in our approach we explicitly distinguish $\mu_b$ and $m_b$
allowing us to include an explicit estimate of
the resummation uncertainty associated with the 5F resummation by varying the (in principle arbitrary) matching
scale $\mub$.
%

\subsection{Higher-order two-loop terms: NLO+NNLL$_\pa$}
\label{subsec:nlonnllp}

At present, all coefficient functions in \eq{bbhCountingRes} required up to NLO+NLL are known~\cite{Harlander:2003ai,Buehler:2012cu,Dittmaier:2003ej,Dawson:2003kb}.
Going to NNLO+NNLL is not yet possible and would require the full NNLO 4FS result, corresponding to the unknown
$\bar C^{(4)}_{ij}$ and $C^{(3)}_{bk}$ coefficients in \eq{bbhCountingRes},
together with the three-loop matching coefficients $\mathcal{M}_{ij}$.
The two-loop coefficients $C^{(2)}_{b\bar b}$ and $C^{(2)}_{bb}$ are known from the NNLO 5FS
result~\cite{Harlander:2003ai} and are part of the NNLL resummation in our counting. Adding them to the NLO+NLL result
provides a partial NNLL result, see \fig{bbh-diagrams}, which we denote as NLO+NNLL$_\pa$.
(In \mycite{Bonvini:2015pxa}, this result was called NLO+NLL+$C^{(2)}_{b\bar b}$.
It corresponds to what in \mycite{Forte:2015hba} would be called FONLL-B.)

Including the partial NNLL terms violates the exact correspondence between the resummed and fixed-order results.
That is, the fixed-order limit of these resummed terms are only reproduced by the $m_b\to 0$ limit of
the fixed-order result at NNLO. Hence, in the limit $\mu_b\to \mu_F$ these terms spoil the smooth matching of the
matched result into the fixed-order result. In this regard, these terms are analogous to the higher-order cross
terms that are kept when the strict expansion is not implemented.
Therefore, whenever the strict expansion is used these terms should also be consistently dropped.
This is the case when going toward larger values of $m_b/m_H$, where the logarithms become small and fixed-order contributions become dominant, and the matched result is transitioning into the pure fixed-order result.

For intermediate values of $m_H$ (small values of $m_b/m_H$) where our perturbative counting is most appropriate,
including these terms does not necessarily improve the overall accuracy of the prediction, since other terms of the same
perturbative order are not included and including only a partial set of terms might bias the result in the wrong direction.
At the same time, their inclusion can lead to a reduced scale dependence, which would then potentially underestimate the
perturbative uncertainties.
For these reasons, we do not take the NLO+NNLL$_\pa$ prediction to be our default result. Nevertheless,
we also provide it in \sec{lhc13}, as it can provide an indication of the numerical size of the next-order correction.
This can for example be an additional useful cross check on the estimate of the
perturbative uncertainties of the NLO+NLL result or alternatively guide the choice of the central scales.

Finally, in the limit of very large $m_H$ (very small values of $m_b/m_H$), including these terms becomes beneficial once the size of the resummed logarithms grows, $\as\ln(\muf^2/\mub^2)\sim 1$, and the original strict 5FS counting applies.

\subsection{Fixed-order nonsingular $Y_b Y_t$ contributions}
\label{subsec:ybyt}

\begin{figure}[t]
\centering
\includegraphics[trim=2.1cm 18.6cm 1.6cm 4.3cm,clip,width=0.7\textwidth]{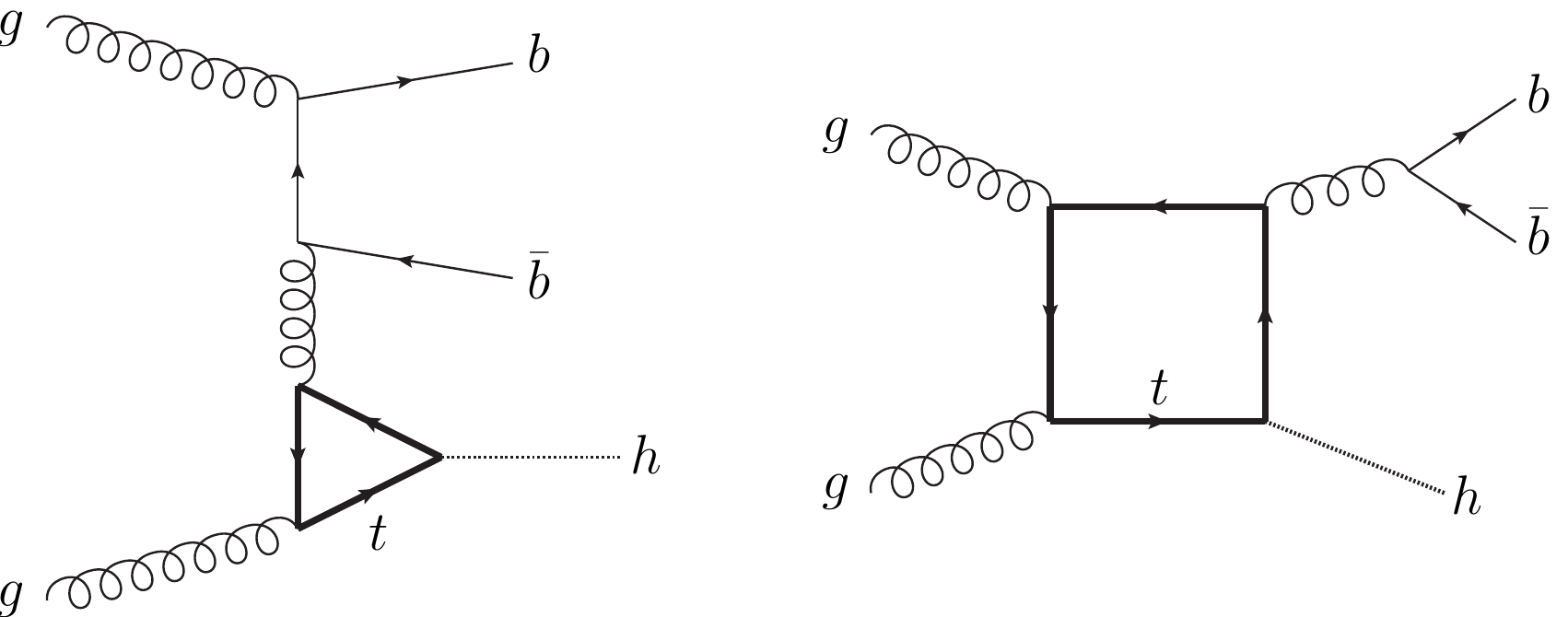}
\caption{Sample 1-loop diagrams contributing to the $Y_bY_t$ interference contribution at fixed order.}
\label{fig:ybyt-diagrams}
\end{figure}

At LO there is only a contribution proportional to $Y^2_b$.
Starting at NLO, the cross section receives contributions proportional to $Y_b Y_t$ due to the
interference of the Born-level $gg\to b\bar b H$ diagrams with diagrams where the Higgs is radiated from a closed top-quark loop;
some examples of the latter diagrams are shown in \fig{ybyt-diagrams}.
The fixed-order (4FS) cross section can be written schematically as
\begin{align} \label{eq:FOybyt}
\sigma^{\rm FO} &= \as^2\, Y_b^2\, \sigma^{(0)} + \as^3 \Bigl( Y_b^2\, \sigma^{(1)}_{Y_b^2} +  Y_b Y_t\, \sigma^{(1)}_{Y_b Y_t} \Bigr)
+ \Ord(\as^4)
\,,\end{align} 
where the interference terms are included in $\sigma^{(1)}_{Y_b Y_t}$.

These top-loop diagrams have the same structure as the contributions that enter the
$gg\to H$ gluon-fusion cross section, so the $Y_b Y_t$ interference contributions fundamentally
correspond to an interference between the $b\bar bH$ and gluon-fusion processes.
Since they involve $b$-quarks in the final state, they are usually regarded as part of the
$b\bar b H$ cross section.

Curiously, these terms can be treated as purely nonsingular terms, even though
they may, at first sight, appear to contain large logarithms of $m_b/m_H$ in the $m_b\to 0$ limit.
On closer inspection, these interference terms turn out to vanish to all orders in the $m_b\to 0$.
The reason is that in the interference between $b\bar{b}H$-like and gluon-fusion-like
diagrams the Higgs boson must be attached to two different closed fermion lines.
This requires a helicity flip on the $b$-quark line, which is not allowed for $m_b=0$. Equivalently,
for $m_b = 0$, they will always contains a trace over an odd number of Dirac matrices and thus vanish~\cite{Harlander:2003ai}.
For the same reason, such terms are absent in the 5FS.

For us, this means that these terms are purely nonsingular and can be straightforwardly
added by including them in $\sigma^{\rm FO}$ in \eq{FOybyt}, which then enters into the $\bar C^{(3)}_{ij}$ in \eq{bbhCountingRes}.
In practice, we extract the numerical result for $\sigma^{(1)}_{Y_b^2}$ and $\sigma^{(1)}_{Y_b Y_t}$ in the pole scheme
from \texttt{Madgraph5\_aMC@NLO}~\cite{Alwall:2014hca} by generating the process $p p \to b \bar{b} H$ at NLO with $Y_t$ turned
on and off. These are then used to construct $\bar C^{(3)}_{ij}$ in the $\MSb$ scheme for the Yukawa couplings.

The $Y_bY_t$ interference terms have a noticeable numerical effect ($\sim5\%$) in the SM,
while in beyond-the-Standard-Model (BSM) scenarios such as SUSY with large $\tan\beta$ their relative
effect compared to the dominant $Y_b^2$ contribution tends to be much milder.
In \sec{lhc13} we therefore provide the results both with and without the $Y_b Y_t$ terms included,
which we denote as \NLO\ and \NLOt\ in the following.
For our choice of central scales, the $Y_b Y_t$ terms reduce the cross section for
$m_H\lesssim300$~GeV and increase it for $m_H\gtrsim300$~GeV, see \fig{mhscan} below.

\section{Estimate of perturbative and parametric uncertainties}
\label{sec:uncertainty}

We now turn to the discussion of the theoretical and parametric uncertainties.
The estimation of the perturbative uncertainties by variations of the hard scales $\muf$ and $\mur$, and low matching scale
$\mub$ are discussed in \subsec{scales}.
The parametric uncertainty from the value of the $b$-quark mass is discussed in \subsec{mberror}.
In \subsec{pdf}, we discuss the PDF uncertainty and the construction of
modified PDF sets that are required to properly separate the $m_b$ and $\mu_b$ uncertainties.

\subsection{Scale choices and perturbative uncertainties}
\label{subsec:scales}

As discussed in detail in \mycite{Bonvini:2015pxa}, we can distinguish two different
sources of perturbative uncertainties. One is an overall ``fixed-order'' uncertainty
(within the resummed or matched results), which can be estimated by exploiting the dependence
on the hard matching scale. The second is a ``resummation'' uncertainty related to the
uncertainty in the resummed logarithmic series, which can be estimated by exploiting the
dependence on the low $\mu_b$ matching scale.

In \mycite{Bonvini:2015pxa} we considered a common hard scale.
Here, we additionally study the dependence of the cross section on the renormalization scale $\mur$
at which $\as$ is evaluated and on the renormalization scale $\mu_Y$ at which the Yukawa coupling is evaluated.
In this case, the role of the hard matching scale in the resummation is played by the factorization scale $\muf$
at which the PDFs are evaluated.

\subsubsection{Central scale choices}

For the factorization scale $\muf$ we use the central choice $\muf=(m_H+2m_b)/4$, as is commonly used in both the 4FS and 5FS
calculations. This choice is motivated by the well-known observation that in $b\bar bH$ such a small factorization scale leads to an
improved perturbative convergence, see e.g.~\cite{Maltoni:2003pn, Maltoni:2005wd, Maltoni:2012pa, Bonvini:2015pxa, Harlander:2015xur}. 
We point out that the matched NLO+NLL result turns out to be significantly less sensitive to the central value of $\muf$~\cite{Bonvini:2015pxa} than the 4FS and 5FS results.
The value of $m_b$ in the definition of $\muf$ is taken to be the central pole mass value $m_b=4.58$~GeV (see \subsec{mberror}),
and is kept fixed under $m_b$ variations.

For the renormalization scale we use a somewhat larger central value $\mur=m_H/2$.
This is motivated by the fact that kinematical arguments for a small scale $\sim m_H/4$ are related
to the collinear factorization ($\muf$) and not the renormalization ($\mur$).
On the other hand, choosing $\mur=m_H$, which would be the canonical renormalization scale, produces somewhat artificial leftover $\ln(\mur/\muf)\simeq\ln 4$ terms in the cross section, which become even larger under scale variations.
The value $\mur=m_H/2$ is a reasonable compromise that lies halfway between the standard 4FS and 5FS choices of
$\mur=(m_H+2m_b)/4$ and $\mur=m_H$. Also, at the Higgs masses of interest, the matched result is dominated by the resummation
contributions and as shown in \mycite{Harlander:2015xur} the 5FS tends to favor $\mur$ values
between $m_H/2$ and $m_H$. Finally, this choice has the convenient side effect that the NLO+NNLL$_\pa$ result turns out to be
a very small correction over the NLO+NLL result.

\begin{figure}[t]
\centering
\includegraphics[scale=0.66]{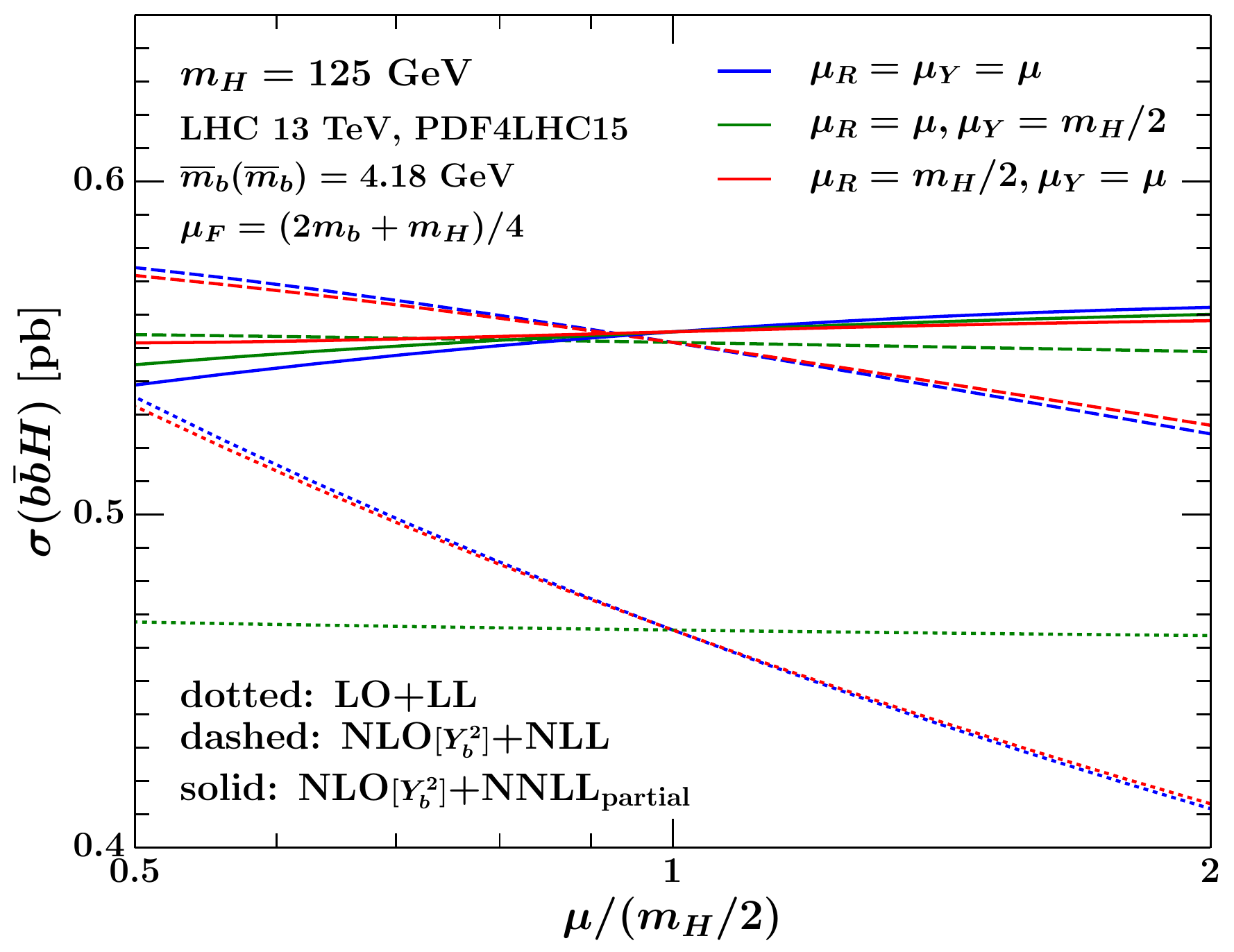}
\caption{Dependence of the cross sections on $\mur$ and $\mu_Y$ at LO$+$LL (dotted),
  \NLO$+$NLL (dashed), and \NLO$+$NNLL$_\pa$ (solid) for $m_H=125$~GeV and 13~TeV.
  The blue curves show the total scale dependence when setting $\mur=\mu_Y=\mu$, the green curves show the dependence
  on $\mur=\mu$ for fixed $\mu_Y=m_H/2$, and the red curves show the dependence on $\mu_Y=\mu$ for fixed $\mur=m_H/2$.
  In call cases $\muf$ is held fixed at its central value.}
\label{fig:murmuy}
\end{figure}

The Yukawa couplings $Y_b(\mu_Y)$ and $Y_t(\mu_Y)$ are defined in the $\MSb$ scheme are obtained by
evolving from $\overline{m}_b(\overline{m}_b)=4.18\GeV$~\cite{Agashe:2014kda}
and $\overline{m}_t(\overline{m}_t)=162.7\GeV$~\cite{Denner:2047636} to the central Yukawa scale $\mu_Y$ with 4-loop evolution, while $\mu_Y$ variations are computed using 2-loop evolution.
While both $\mur$ and $\mu_Y$ are renormalization scales, they do not necessarily need to be the same.
It is always possible to evolve $\as$ and the Yukawa coupling to different scales using their own
renormalization group evolution, compensated by including the appropriate fixed-order logarithms in the partonic coefficients.
In \fig{murmuy} we study the dependence of the cross section on $\mur$ and $\mu_Y$ at different orders, always keeping $\muf$ fixed
at its central value. We find that varying $\mur$ and $\mu_Y$ together gives the largest scale variation, so
for our numerical results and uncertainty estimation we identify $\mu_Y \equiv \mur$ as is usually done.
We also observe that at LO+LL and \NLO+NLL the renormalization scale dependence comes entirely from the
$\mu_Y$ dependence, which reduces significantly at each higher order.
(This is also another motivation to choose a higher central scale for $\mur$ since $m_H$ is the (only)
relevant scale seen by the $b\bar bH$ vertex.) Note that at \NLO+NNLL$_\pa$
the $\mu_Y$ dependence reduces, which is precisely to be expected from the included two-loop virtual corrections.
On the other hand, the $\mur$ dependence for fixed $\mu_Y$ at \NLO+NNLL$_\pa$ actually increases,
which could well be related to only including a partial set of higher-order terms.

We note that this observed pattern for the $\mu$ dependence is somewhat changed by
the inclusion of the $Y_bY_t$ interference terms.
This is not unexpected, since these terms introduce new LO dependence on both
$\mur$ and $\mu_Y$. However, since their absolute correction to the cross section is small,
we base our discussion of the scale choices on the pattern observed without interference terms.

Finally, for $\mub$ we take the canonical central value $\mub = 4.58\GeV$, which corresponds to the central pole mass value we use, but importantly is kept fixed under $m_b$ variations. The canonical scale choice $\mub = m_b$ is appropriate in the resummation region where $\mub \ll \muf$, which is the case for all Higgs masses we consider in this paper. As explained in \mycite{Bonvini:2015pxa}, for larger values of $m_b/\muf \gtrsim 0.3$ (smaller $m_H$) one enters the transition region, where the strict expansion should be used and $\mub$ should be chosen via a more general profile scale to allow for a smooth turning off the resummation and transition into the fixed-order limit $m_b \sim \muf$.

\subsubsection{Estimate of perturbative uncertainties}

In \mycite{Bonvini:2015pxa}, $\mur=\mu_Y=\muf$ were all varied together, which already yields an excellent perturbative convergence.
Here we follow an even more conservative approach and also explore independent variations of $\mur$ and $\muf$.
We consider the usual 7-point variation where the two scales are varied independently up and down by factors of two, excluding the two cases where they are both varied in opposite directions.
Whenever varying $\muf$ up or down, the low matching scale $\mub$ is varied up or down by the same factor, such that the ratio $\muf/\mub$ and therefore the resummed logarithms remain fixed. As discussed in \mycite{Bonvini:2015pxa}, this allows us to interpret the hard scale variations as an estimate of the overall fixed-order uncertainty.
The final fixed-order uncertainty is then obtained by the maximal envelope, i.e., we use the absolute value of the largest deviation from the central value as the symmetric uncertainty.
Doing so explicitly avoids attributing any physical meaning to accidentally small one-sided scale dependence (that would yield asymmetric uncertainties), which just results from a nonlinear scale dependence as frequently encountered at higher orders or near the points of minimal scale dependence.

Next, the resummation uncertainty is computed by varying the matching scale $\mub$ by a factor of 2 up and down
about its central value. For this variation we keep all the other scales fixed (and also the value of $m_b$).
Doing so explicitly changes the ratio $\muf/\mub$ and thus directly probes the size of the resummed logarithms.

The fixed-order and resummation uncertainties are considered as independent uncertainty sources, and the total perturbative uncertainty is obtained by adding them in quadrature.
A simple alternative approach, which however lacks the physical interpretation of the source of uncertainty, would be to
consider all possible independent variations of $\muf$, $\mur$, $\mub$ by factors of two, eliminating all cases where the ratio of any two
variation factors exceeds $2$, and taking the total envelope. This turns out to be more aggressive and produces a smaller total uncertainty, because some of the variations (the $\mub$ variations in particular) that in our approach are considered independent and added in quadrature, are simply contained within the overall envelope and thus have no effect on the final uncertainty.

\subsection{Parametric uncertainties due to $m_b$}
\label{subsec:mberror}

We now discuss the settings we use for the $b$-quark mass.
The $b$-quark mass is most precisely measured when defined in a renormalon-free short-distance scheme, such as the $\MSb$ or $1S$ schemes.
As our starting point and central input value we thus take the measured value of the $\MSb$ mass $\overline{m}_b(\overline{m}_b)=4.18 \pm 0.03$~GeV~\cite{Agashe:2014kda}.

In principle, the best option would be to always use a renormalon-free mass renormalization scheme along with the corresponding measured value in all the places where the mass appears. These are the Yukawa coupling $Y_b$, the $m_b$ dependence in the nonsingular parts of the coefficient functions, and the $m_b$ dependence of the PDF matching coefficients $\mathcal{M}_{ij}(m_b, \mu_b)$ [see \eq{bbH5FSexplicit}].
As mentioned in \subsec{scales} above, the Yukawa coupling is renormalized in the $\MSb$ scheme and is directly obtained
by evolving from $\mu = \overline{m}_b$ to $\mur$.

Unfortunately, most current PDF fits are performed with pole-scheme masses, including all PDF sets that currently enter the PDF4LHC15 combination, which is what we will utilize, see \subsec{pdf} below.
Furthermore, the fixed-order computations that we use as input and hence our nonsingular coefficient functions are currently obtained in the pole scheme. To be consistent, we thus also require $m_b$ in the pole scheme.

Since the pole mass has a leading renormalon ambiguity, the choice of its value is somewhat delicate.
To reproduce as closely as possible the renormalon cancellation that would happen when properly translating the perturbative expressions from the pole scheme to the $\MSb$ scheme, we have to convert the $\overline{m}_b$ input value to a pole-mass value at the same loop order at which the perturbative series where $m_b$ appears (and that contains the cancelling renormalon) is used. In our case, this means we should translate it at 1-loop order because the proper NLO+NLL $\MSb$ result would require an NNLO $\MSb$
PDF set and would be affected by the 1-loop pole to $\MSb$ conversion,
\begin{equation}
  \label{eq:mbpole}
  \overline{m}_b(\overline{m}_b)=4.18~\text{GeV} \qquad \overset{\text{1-loop}}{\longrightarrow}\qquad
  m_b^{\text{pole}} = 4.58~\text{GeV}
\,.\end{equation}

To see this, note that the $m_b$ dependence of the fixed-order contributions first appears at LO and we work to NLO, so the scheme translation requires the 1-loop conversion. Equivalently, in the PDFs, the $b$-quark mass enters in the $b$-quark matching coefficient $\mathcal{M}_{bi}(m_b, \mu_b)$, which first appears in the NLO PDFs, while the NNLO PDFs contain its 1-loop correction. Note that here it is again important that in our perturbative counting the fixed-order contributions and their singular limit contained in the resummed contributions are consistently included at the same order.

As a cross check, we have explicitly verified that evolving (with \texttt{APFEL}~\cite{Bertone:2013vaa})
the same initial 4-flavor PDF set at NNLO using either the pole scheme with $m_b^{\text{pole}} = 4.58$~GeV
or the $\MSb$ scheme with $\overline{m}_b(\overline{m}_b)=4.18$~GeV gives indeed very similar results.

To evaluate the parametric uncertainty due to the uncertainty in the measured value of $m_b$, we first note that
the current world average for $\overline{m}_b$ has an uncertainty of $\pm30$~MeV~\cite{Agashe:2014kda}.
Given the current tensions in different extractions of $\overline{m}_b$, this uncertainty might be considered too optimistic and one might want to consider the $2\sigma$ variation. For our purposes however the resulting uncertainty in $Y_b$ is small compared to other uncertainties, and so we will use the $1\sigma$ variation
\begin{equation}
  \label{eq:mbmsbar-variation}
  4.15~\GeV \leq \overline{m}_b(\overline{m}_b) \leq 4.21\GeV
\,,\end{equation}
which is directly translated into the variation of the $\MSb$ Yukawa coupling.

Since the conversion to the pole scheme is performed at 1-loop, we also have to take into account the intrinsic uncertainty in the conversion, which is much larger than the uncertainty on $\overline{m}_b$ itself. The 2-loop conversion yields $4.72\GeV$, and we take the difference of $140$~MeV with respect to the 1-loop conversion as a reasonably conservative estimate which should be sufficient to cover the uncertainties in $\overline{m}_b$
and the conversion to the pole scheme. Therefore, to estimate the parametric uncertainty in $m_b$ in our predictions we use the variation
\begin{equation}  \label{eq:mbpole-variation}
  4.44~\GeV \leq m_b^{\text{pole}} \leq 4.72 \GeV
\,,\end{equation}
where the lower (upper) variations on $m_b^\text{pole}$ are always used in conjunction with the lower (upper) variation on $\overline{m}_b$ in \eq{mbmsbar-variation}. As we will see, the uncertainty due to $m_b$ will be small.

Finally, as mentioned earlier, when varying $m_b$, the $b$-quark matching scale is kept fixed at its central value $\mub=4.58$~GeV and $\muf$ is also kept fixed at its central value $\muf=(m_H+2m_b)/4$ with $m_b=4.58$~GeV.

\subsection{Input PDFs and PDF uncertainties}
\label{subsec:pdf}

As discussed in \sec{nlonll}, the resummation of collinear logarithms is entirely contained in the 
evolved 5FS PDFs, and these carry a physical dependence on $m_b$ and an unphysical
dependence on the $b$-quark matching scale $\mub$.
When computing the cross section, it is important that the input parameters used are
consistent with the used PDF set.
In particular, in our matched predictions we have to ensure that the value of $m_b$ in the computation of the
fixed-order contributions is equal to the one present in the resummed contributions, i.e., in the PDF set, since
otherwise the nonsingular terms would receive residual singular logarithmic terms arising from miscancellations.

To be able to use in a fully consistent way all the values we want for $m_b$ and $\mu_b$, for the central value predictions
as well as the uncertainty estimation, we require dedicated PDFs that are not available by default.
In principle, when changing the internal parameters of the PDFs, they should be refitted.
In practice, the chosen value of $m_b$ when fitting the PDFs has a very small effect for all PDFs except for the
$b$-quark PDF itself~\cite{Harland-Lang:2015qea}. The reason is that the presently fitted data provides only
a very weak direct constraint on the $b$-quark PDF or $m_b$, which means that for all practical purposes the
$b$-quark PDF is essentially being \emph{calculated} from the fitted gluon and light-quark PDFs at the low scale $\mu_0 < \mu_b$.
Hence, we can also safely assume that refitting the PDFs for $\mub \neq m_b$ will not have much effect
on the light-quark and gluon PDFs at $\mu_0$.

Given the above, we can take any PDF set with a given value of $\mu_b = m_b$, and re-evolve it starting
from a low scale $\mu_0 <\mub$ but using different values for $m_b$, $\mu_b$, and the mass renormalization scheme.
In other words, we use the light-quark and gluon PDFs at $\mu_0$ as the input quantity and compute the $b$-quark PDF ourselves.
(Note that this procedure is also typically used by PDF fitting groups when constructing fixed-flavor PDF sets from the fitted variable-flavor sets.) This approach is very useful because it opens the possibility of using any desired
values for $\mu_b$ and $m_b$ with any particular input PDF set.

Regarding the input PDFs at $\mu_0$, we use the combined \texttt{PDF4LHC15\_nnlo\_mc} set~\cite{Butterworth:2015oua, Carrazza:2015hva,Ball:2014uwa,Harland-Lang:2014zoa,Dulat:2015mca}.
All 101 PDF members are re-evolved from an initial scale $\mu_0 = 2\GeV$ for all the values of
$m_b$ and $\mu_b$ that we need.
This is done using the latest version ($\geq 2.8$) of \texttt{APFEL}~\cite{Bertone:2013vaa}.%
\footnote{In \mycite{Bonvini:2015pxa} we had used a privately modified version of \texttt{APFEL} to allow for
heavy-quark thresholds that are different from the mass, $\mub \neq m_b$. This feature is now available in
the latest version of \texttt{APFEL}.}
We emphasize that re-evolving the PDF4LHC15 PDFs in this way is in fact more consistent for the case of $b\bar bH$ than
directly using the $b$-quark PDF from the combined set, because
the prior sets (MMHT2014~\cite{Harland-Lang:2014zoa}, CT14~\cite{Dulat:2015mca}, NNPDF30~\cite{Ball:2014uwa})
from which the PDF4LHC15 set is constructed use different values of $m_b$.
The re-evolved PDF sets for various different values of $\mu_b$ and $m_b$ are publicly available at
\href{http://www.ge.infn.it/~bonvini/bbh}{\texttt{http://www.ge.infn.it/$\sim$bonvini/bbh}}.

To compute the PDF uncertainties we follow the PDF4LHC prescription~\cite{Butterworth:2015oua}, using our re-evolved set
with our central values for $m_b$ and $\mu_b$.
That is, we compute the cross section for all 100 replicas, order the results in ascending order, and obtain a 68\% confidence level interval by using the 17th result as the lower variation and the 84th result as the upper variation.
We note that the distribution of cross sections we obtain is Gaussian to a very good approximation.

\section{Results for the 13 TeV LHC}
\label{sec:lhc13}

\begin{table}[tp]
\renewcommand{\arraystretch}{1.1}
\centering
\begin{tabular}{c|ccccc}
\hline\hline
\multicolumn{6}{c}{\NLO+NLL}
\\\hline
  $m_H$ [GeV] & $\sigma(b\bar b H)$ [pb] &   $\{\mur,\muf\}$ [\%] &   $\mub$ [\%] &     $m_b$ [\%] &    PDFs [\%]
\\\hline
  50 & $7.46\times10^{+0}$ & $\pm20.0$ & $\pm7.2$ & $\pm3.0$ & ${}^{+1.5}_{-3.6}$ \\
  75 & $2.57\times10^{+0}$ & $\pm12.6$ & $\pm5.7$ & $\pm2.3$ & ${}^{+1.7}_{-3.0}$ \\
 100 & $1.11\times10^{+0}$ & $\pm8.7$ & $\pm5.5$ & $\pm1.9$ & ${}^{+2.0}_{-2.6}$ \\
 125 & $5.52\times10^{-1}$ & $\pm8.9$ & $\pm5.3$ & $\pm1.7$ & ${}^{+2.1}_{-2.8}$ \\
 150 & $3.02\times10^{-1}$ & $\pm9.0$ & $\pm5.1$ & $\pm1.7$ & ${}^{+2.6}_{-2.4}$ \\
 175 & $1.78\times10^{-1}$ & $\pm9.1$ & $\pm5.0$ & $\pm1.4$ & ${}^{+2.2}_{-2.7}$ \\
 200 & $1.11\times10^{-1}$ & $\pm9.2$ & $\pm5.0$ & $\pm1.4$ & ${}^{+2.1}_{-2.5}$ \\
 225 & $7.23\times10^{-2}$ & $\pm9.2$ & $\pm4.9$ & $\pm1.2$ & ${}^{+2.5}_{-2.6}$ \\
 250 & $4.87\times10^{-2}$ & $\pm9.3$ & $\pm5.0$ & $\pm1.3$ & ${}^{+2.4}_{-2.6}$ \\
 275 & $3.38\times10^{-2}$ & $\pm9.2$ & $\pm4.8$ & $\pm1.1$ & ${}^{+2.6}_{-2.6}$ \\
 300 & $2.40\times10^{-2}$ & $\pm9.2$ & $\pm4.7$ & $\pm1.2$ & ${}^{+2.7}_{-2.9}$ \\
 325 & $1.75\times10^{-2}$ & $\pm9.4$ & $\pm4.9$ & $\pm1.3$ & ${}^{+2.2}_{-3.1}$ \\
 350 & $1.29\times10^{-2}$ & $\pm9.4$ & $\pm4.8$ & $\pm1.2$ & ${}^{+2.3}_{-3.3}$ \\
 375 & $9.67\times10^{-3}$ & $\pm9.1$ & $\pm4.6$ & $\pm1.2$ & ${}^{+2.4}_{-3.3}$ \\
 400 & $7.38\times10^{-3}$ & $\pm9.2$ & $\pm4.4$ & $\pm1.1$ & ${}^{+3.0}_{-3.3}$ \\
 425 & $5.71\times10^{-3}$ & $\pm9.5$ & $\pm4.8$ & $\pm0.9$ & ${}^{+2.7}_{-3.4}$ \\
 450 & $4.45\times10^{-3}$ & $\pm9.4$ & $\pm4.7$ & $\pm0.9$ & ${}^{+3.0}_{-3.6}$ \\
 475 & $3.51\times10^{-3}$ & $\pm9.4$ & $\pm4.6$ & $\pm0.9$ & ${}^{+3.6}_{-3.6}$ \\
 500 & $2.80\times10^{-3}$ & $\pm9.7$ & $\pm5.1$ & $\pm1.1$ & ${}^{+3.9}_{-3.7}$ \\
 525 & $2.24\times10^{-3}$ & $\pm9.6$ & $\pm4.9$ & $\pm0.9$ & ${}^{+4.3}_{-3.8}$ \\
 550 & $1.81\times10^{-3}$ & $\pm9.4$ & $\pm4.7$ & $\pm0.9$ & ${}^{+4.4}_{-3.8}$ \\
 600 & $1.21\times10^{-3}$ & $\pm9.5$ & $\pm4.6$ & $\pm0.9$ & ${}^{+5.3}_{-3.8}$ \\
 650 & $8.25\times10^{-4}$ & $\pm9.8$ & $\pm5.1$ & $\pm0.9$ & ${}^{+7.0}_{-3.5}$ \\
 700 & $5.75\times10^{-4}$ & $\pm9.7$ & $\pm4.8$ & $\pm0.9$ & ${}^{+6.2}_{-3.9}$ \\
 750 & $4.08\times10^{-4}$ & $\pm9.7$ & $\pm4.8$ & $\pm0.8$ & ${}^{+7.5}_{-4.1}$ \\
\hline\hline
\end{tabular}
\caption{\NLO+NLL cross section predictions for the LHC at 13~TeV with individual error estimates for 
$\{\muf,\mur\}$, $\mub$, $m_b$ and PDF uncertainties. See text for further details.}
\label{tab:nlonll}
\end{table}

\begin{table}[tp]
\renewcommand{\arraystretch}{1.1}
\centering
\begin{tabular}{c|ccccc}
\hline\hline
\multicolumn{6}{c}{\NLOt+NLL}
\\\hline
  $m_H$ [GeV] & $\sigma(b\bar b H)$ [pb] &   $\{\mur,\muf\}$ [\%] &   $\mub$ [\%] &     $m_b$ [\%] &    PDFs [\%]
\\\hline
  50 & $7.00\times10^{+0}$ & $\pm20.0$ & $\pm7.7$ & $\pm3.0$ & ${}^{+1.6}_{-3.8}$ \\
  75 & $2.42\times10^{+0}$ & $\pm12.8$ & $\pm5.9$ & $\pm2.6$ & ${}^{+1.8}_{-3.2}$ \\
 100 & $1.05\times10^{+0}$ & $\pm8.9$ & $\pm6.0$ & $\pm2.1$ & ${}^{+2.2}_{-2.8}$ \\
 125 & $5.25\times10^{-1}$ & $\pm7.7$ & $\pm5.8$ & $\pm1.9$ & ${}^{+2.2}_{-3.0}$ \\
 150 & $2.89\times10^{-1}$ & $\pm7.7$ & $\pm5.5$ & $\pm1.6$ & ${}^{+2.7}_{-2.5}$ \\
 175 & $1.72\times10^{-1}$ & $\pm8.0$ & $\pm5.3$ & $\pm1.7$ & ${}^{+2.3}_{-2.8}$ \\
 200 & $1.08\times10^{-1}$ & $\pm8.2$ & $\pm4.9$ & $\pm1.4$ & ${}^{+2.1}_{-2.6}$ \\
 225 & $7.07\times10^{-2}$ & $\pm8.4$ & $\pm5.1$ & $\pm1.3$ & ${}^{+2.6}_{-2.6}$ \\
 250 & $4.79\times10^{-2}$ & $\pm8.6$ & $\pm4.4$ & $\pm1.4$ & ${}^{+2.5}_{-2.7}$ \\
 275 & $3.35\times10^{-2}$ & $\pm8.9$ & $\pm4.9$ & $\pm1.3$ & ${}^{+2.6}_{-2.6}$ \\
 300 & $2.40\times10^{-2}$ & $\pm8.9$ & $\pm4.8$ & $\pm1.6$ & ${}^{+2.7}_{-2.9}$ \\
 325 & $1.77\times10^{-2}$ & $\pm9.4$ & $\pm4.8$ & $\pm1.5$ & ${}^{+2.2}_{-3.1}$ \\
 350 & $1.32\times10^{-2}$ & $\pm9.7$ & $\pm4.3$ & $\pm1.9$ & ${}^{+2.2}_{-3.2}$ \\
 375 & $1.00\times10^{-2}$ & $\pm9.9$ & $\pm4.1$ & $\pm1.9$ & ${}^{+2.3}_{-3.2}$ \\
 400 & $7.79\times10^{-3}$ & $\pm10.3$ & $\pm4.1$ & $\pm1.2$ & ${}^{+2.8}_{-3.1}$ \\
 425 & $6.02\times10^{-3}$ & $\pm10.5$ & $\pm4.7$ & $\pm1.1$ & ${}^{+2.5}_{-3.2}$ \\
 450 & $4.72\times10^{-3}$ & $\pm10.7$ & $\pm4.1$ & $\pm1.0$ & ${}^{+2.9}_{-3.4}$ \\
 475 & $3.72\times10^{-3}$ & $\pm10.5$ & $\pm4.1$ & $\pm1.1$ & ${}^{+3.4}_{-3.4}$ \\
 500 & $2.96\times10^{-3}$ & $\pm11.0$ & $\pm5.0$ & $\pm1.2$ & ${}^{+3.7}_{-3.5}$ \\
 525 & $2.37\times10^{-3}$ & $\pm11.0$ & $\pm4.9$ & $\pm1.1$ & ${}^{+4.0}_{-3.6}$ \\
 550 & $1.91\times10^{-3}$ & $\pm10.8$ & $\pm4.6$ & $\pm0.9$ & ${}^{+4.2}_{-3.6}$ \\
 600 & $1.26\times10^{-3}$ & $\pm10.4$ & $\pm4.2$ & $\pm1.3$ & ${}^{+5.1}_{-3.6}$ \\
 650 & $8.58\times10^{-4}$ & $\pm10.6$ & $\pm5.0$ & $\pm1.3$ & ${}^{+6.7}_{-3.4}$ \\
 700 & $5.96\times10^{-4}$ & $\pm10.6$ & $\pm4.8$ & $\pm1.0$ & ${}^{+6.0}_{-3.8}$ \\
 750 & $4.20\times10^{-4}$ & $\pm10.4$ & $\pm4.9$ & $\pm1.0$ & ${}^{+7.3}_{-4.0}$ \\
\hline\hline
\end{tabular}
\caption{\NLOt+NLL cross section predictions for the LHC at 13~TeV with individual error estimates for
$\{\muf,\mur\}$, $\mub$, $m_b$ and PDF uncertainties. See text for further details.}
\label{tab:nlonllybyt}
\end{table}

\begin{table}[tp]
\renewcommand{\arraystretch}{1.1}
\centering
\begin{tabular}{c|ccccc}
\hline\hline
\multicolumn{6}{c}{\NLO+NNLL$_\pa$}
\\\hline
  $m_H$ [GeV] & $\sigma(b\bar b H)$ [pb] &   $\{\mur,\muf\}$ [\%] &   $\mub$ [\%] &     $m_b$ [\%] &    PDFs [\%]
\\\hline
  50 & $7.53\times10^{+0}$ & $\pm20.7$ & $\pm7.2$ & $\pm3.0$ & ${}^{+1.5}_{-3.5}$ \\
  75 & $2.59\times10^{+0}$ & $\pm13.2$ & $\pm5.7$ & $\pm2.3$ & ${}^{+1.7}_{-3.0}$ \\
 100 & $1.12\times10^{+0}$ & $\pm9.1$ & $\pm5.5$ & $\pm1.9$ & ${}^{+2.0}_{-2.6}$ \\
 125 & $5.55\times10^{-1}$ & $\pm6.9$ & $\pm5.3$ & $\pm1.7$ & ${}^{+2.1}_{-2.8}$ \\
 150 & $3.04\times10^{-1}$ & $\pm5.8$ & $\pm5.1$ & $\pm1.7$ & ${}^{+2.6}_{-2.4}$ \\
 175 & $1.79\times10^{-1}$ & $\pm4.5$ & $\pm5.0$ & $\pm1.4$ & ${}^{+2.2}_{-2.7}$ \\
 200 & $1.12\times10^{-1}$ & $\pm3.8$ & $\pm5.0$ & $\pm1.4$ & ${}^{+2.1}_{-2.5}$ \\
 225 & $7.26\times10^{-2}$ & $\pm3.4$ & $\pm4.9$ & $\pm1.2$ & ${}^{+2.5}_{-2.6}$ \\
 250 & $4.89\times10^{-2}$ & $\pm3.2$ & $\pm5.0$ & $\pm1.3$ & ${}^{+2.4}_{-2.6}$ \\
 275 & $3.39\times10^{-2}$ & $\pm3.2$ & $\pm4.8$ & $\pm1.1$ & ${}^{+2.6}_{-2.6}$ \\
 300 & $2.41\times10^{-2}$ & $\pm3.2$ & $\pm4.7$ & $\pm1.2$ & ${}^{+2.7}_{-2.9}$ \\
 325 & $1.75\times10^{-2}$ & $\pm3.4$ & $\pm4.9$ & $\pm1.3$ & ${}^{+2.2}_{-3.1}$ \\
 350 & $1.29\times10^{-2}$ & $\pm3.4$ & $\pm4.8$ & $\pm1.2$ & ${}^{+2.3}_{-3.3}$ \\
 375 & $9.70\times10^{-3}$ & $\pm3.2$ & $\pm4.6$ & $\pm1.2$ & ${}^{+2.4}_{-3.3}$ \\
 400 & $7.41\times10^{-3}$ & $\pm3.2$ & $\pm4.4$ & $\pm1.1$ & ${}^{+3.0}_{-3.3}$ \\
 425 & $5.72\times10^{-3}$ & $\pm3.6$ & $\pm4.8$ & $\pm0.9$ & ${}^{+2.7}_{-3.4}$ \\
 450 & $4.46\times10^{-3}$ & $\pm3.4$ & $\pm4.7$ & $\pm0.9$ & ${}^{+3.0}_{-3.6}$ \\
 475 & $3.52\times10^{-3}$ & $\pm3.4$ & $\pm4.6$ & $\pm0.9$ & ${}^{+3.6}_{-3.6}$ \\
 500 & $2.81\times10^{-3}$ & $\pm3.9$ & $\pm5.1$ & $\pm1.1$ & ${}^{+3.9}_{-3.7}$ \\
 525 & $2.25\times10^{-3}$ & $\pm3.8$ & $\pm4.9$ & $\pm0.9$ & ${}^{+4.1}_{-3.9}$ \\
 550 & $1.81\times10^{-3}$ & $\pm3.6$ & $\pm4.7$ & $\pm0.9$ & ${}^{+4.4}_{-3.8}$ \\
 600 & $1.21\times10^{-3}$ & $\pm3.6$ & $\pm4.6$ & $\pm0.9$ & ${}^{+5.4}_{-3.6}$ \\
 650 & $8.28\times10^{-4}$ & $\pm4.0$ & $\pm5.1$ & $\pm0.9$ & ${}^{+7.0}_{-3.5}$ \\
 700 & $5.77\times10^{-4}$ & $\pm3.8$ & $\pm4.8$ & $\pm0.9$ & ${}^{+6.2}_{-3.9}$ \\
 750 & $4.10\times10^{-4}$ & $\pm3.9$ & $\pm4.8$ & $\pm0.8$ & ${}^{+7.5}_{-4.1}$ \\
\hline\hline
\end{tabular}
\caption{\NLO+NNLL$_\pa$ cross section predictions for the LHC at 13~TeV with individual error estimates for
$\{\muf,\mur\}$, $\mub$, $m_b$ and PDF uncertainties. See text for further details.}
\label{tab:nlonnllp}
\end{table}

\begin{table}[tp]
\renewcommand{\arraystretch}{1.1}
\centering
\begin{tabular}{c|ccccc}
\hline\hline
\multicolumn{6}{c}{\NLOt+NNLL$_\pa$}
\\\hline
  $m_H$ [GeV] & $\sigma(b\bar b H)$ [pb] &   $\{\mur,\muf\}$ [\%] &   $\mub$ [\%] &     $m_b$ [\%] &    PDFs [\%]
\\\hline
  50 & $7.08\times10^{+0}$ & $\pm20.7$ & $\pm7.7$ & $\pm3.0$ & ${}^{+1.6}_{-3.7}$ \\
  75 & $2.44\times10^{+0}$ & $\pm13.4$ & $\pm5.9$ & $\pm2.6$ & ${}^{+1.8}_{-3.2}$ \\
 100 & $1.06\times10^{+0}$ & $\pm9.4$ & $\pm6.0$ & $\pm2.1$ & ${}^{+2.1}_{-2.8}$ \\
 125 & $5.29\times10^{-1}$ & $\pm7.5$ & $\pm5.8$ & $\pm1.9$ & ${}^{+2.2}_{-3.0}$ \\
 150 & $2.91\times10^{-1}$ & $\pm7.1$ & $\pm5.5$ & $\pm1.6$ & ${}^{+2.7}_{-2.5}$ \\
 175 & $1.73\times10^{-1}$ & $\pm6.1$ & $\pm5.3$ & $\pm1.7$ & ${}^{+2.3}_{-2.8}$ \\
 200 & $1.08\times10^{-1}$ & $\pm5.7$ & $\pm4.9$ & $\pm1.4$ & ${}^{+2.1}_{-2.6}$ \\
 225 & $7.10\times10^{-2}$ & $\pm5.5$ & $\pm5.1$ & $\pm1.3$ & ${}^{+2.5}_{-2.6}$ \\
 250 & $4.81\times10^{-2}$ & $\pm4.3$ & $\pm4.4$ & $\pm1.4$ & ${}^{+2.5}_{-2.7}$ \\
 275 & $3.37\times10^{-2}$ & $\pm3.9$ & $\pm4.9$ & $\pm1.3$ & ${}^{+2.6}_{-2.6}$ \\
 300 & $2.41\times10^{-2}$ & $\pm3.2$ & $\pm4.8$ & $\pm1.6$ & ${}^{+2.7}_{-2.9}$ \\
 325 & $1.78\times10^{-2}$ & $\pm2.8$ & $\pm4.8$ & $\pm1.5$ & ${}^{+2.2}_{-3.1}$ \\
 350 & $1.33\times10^{-2}$ & $\pm2.9$ & $\pm4.3$ & $\pm1.9$ & ${}^{+2.2}_{-3.2}$ \\
 375 & $1.01\times10^{-2}$ & $\pm2.9$ & $\pm4.1$ & $\pm1.9$ & ${}^{+2.3}_{-3.2}$ \\
 400 & $7.81\times10^{-3}$ & $\pm3.1$ & $\pm4.1$ & $\pm1.2$ & ${}^{+2.8}_{-3.1}$ \\
 425 & $6.03\times10^{-3}$ & $\pm3.4$ & $\pm4.7$ & $\pm1.1$ & ${}^{+2.5}_{-3.2}$ \\
 450 & $4.74\times10^{-3}$ & $\pm3.5$ & $\pm4.1$ & $\pm1.0$ & ${}^{+2.9}_{-3.4}$ \\
 475 & $3.73\times10^{-3}$ & $\pm3.3$ & $\pm4.1$ & $\pm1.1$ & ${}^{+3.4}_{-3.4}$ \\
 500 & $2.97\times10^{-3}$ & $\pm4.0$ & $\pm5.0$ & $\pm1.2$ & ${}^{+3.7}_{-3.5}$ \\
 525 & $2.38\times10^{-3}$ & $\pm3.9$ & $\pm4.9$ & $\pm1.1$ & ${}^{+3.9}_{-3.7}$ \\
 550 & $1.92\times10^{-3}$ & $\pm3.8$ & $\pm4.6$ & $\pm0.9$ & ${}^{+4.2}_{-3.6}$ \\
 600 & $1.27\times10^{-3}$ & $\pm3.4$ & $\pm4.2$ & $\pm1.3$ & ${}^{+5.2}_{-3.5}$ \\
 650 & $8.61\times10^{-4}$ & $\pm3.9$ & $\pm5.0$ & $\pm1.3$ & ${}^{+6.7}_{-3.4}$ \\
 700 & $5.98\times10^{-4}$ & $\pm4.0$ & $\pm4.8$ & $\pm1.0$ & ${}^{+5.9}_{-3.8}$ \\
 750 & $4.22\times10^{-4}$ & $\pm4.0$ & $\pm4.9$ & $\pm1.0$ & ${}^{+7.3}_{-4.0}$ \\
\hline\hline
\end{tabular}
\caption{\NLOt+NNLL$_\pa$ cross section predictions for the LHC at 13~TeV with individual error estimates for
$\{\muf,\mur\}$, $\mub$, $m_b$ and PDF uncertainties. See text for further details.}
\label{tab:nlonnllpybyt}
\end{table}

\begin{figure}[t]
\centering
\includegraphics[scale=0.66]{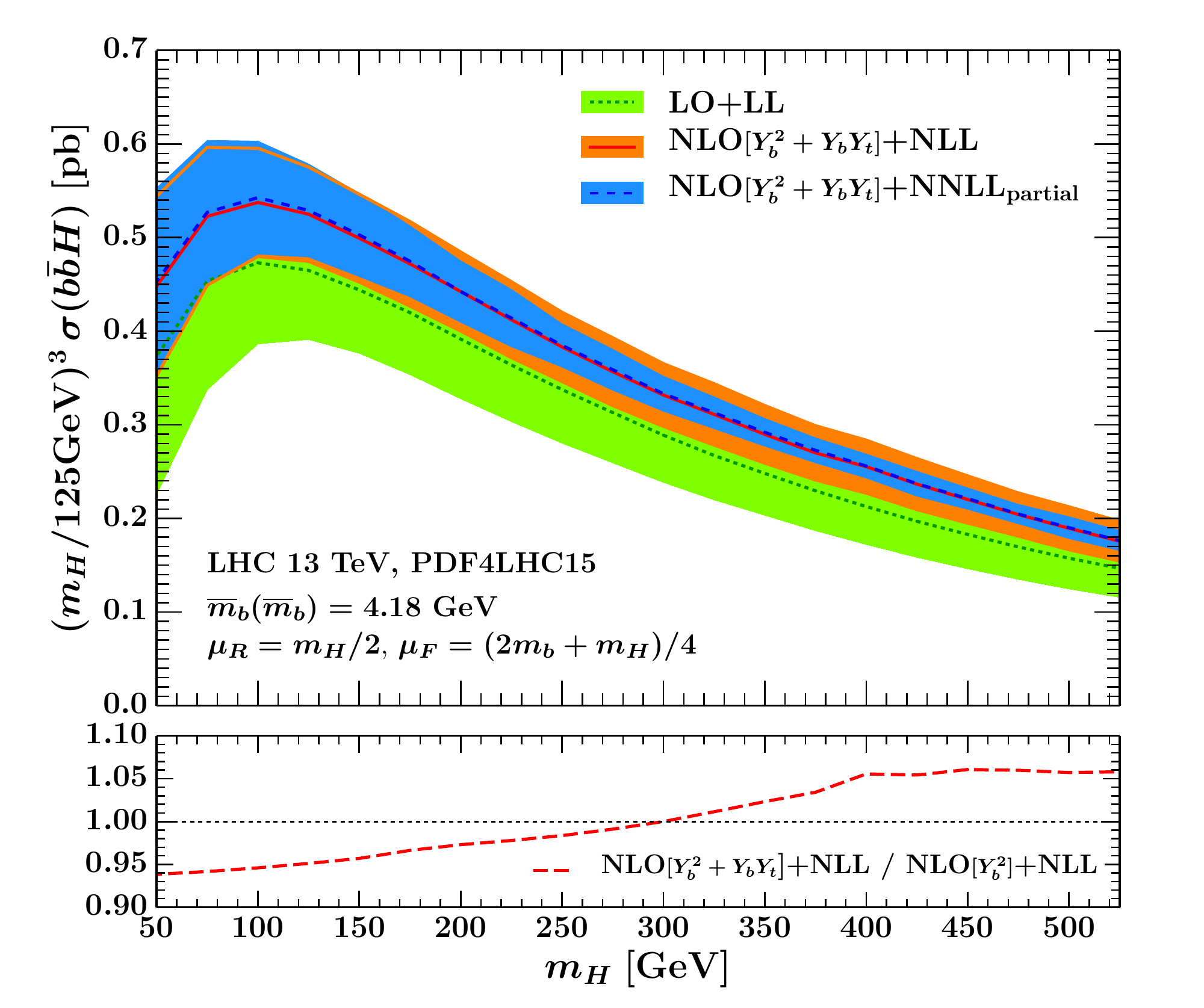}
\caption{Matched $b\bar bH$ cross section as a function of $\mh$, comparing different orders
at LO+LL (green band), \NLOt+NLL (orange band), and \NLOt+NNLL$_\pa$ (blue band). The cross
section is rescaled by $(\mh/125\GeV)^3$.
The lower panel shows the ratio of the central predictions \NLOt+NLL over \NLO+NLL.
The uncertainty bands are obtained by adding the $\{ \mu_F,\mu_R \}$ and $\mu_b$ uncertainties in quadrature.}
\label{fig:mhscan}
\end{figure}

\begin{figure}[t]
\centering
\includegraphics[scale=0.75]{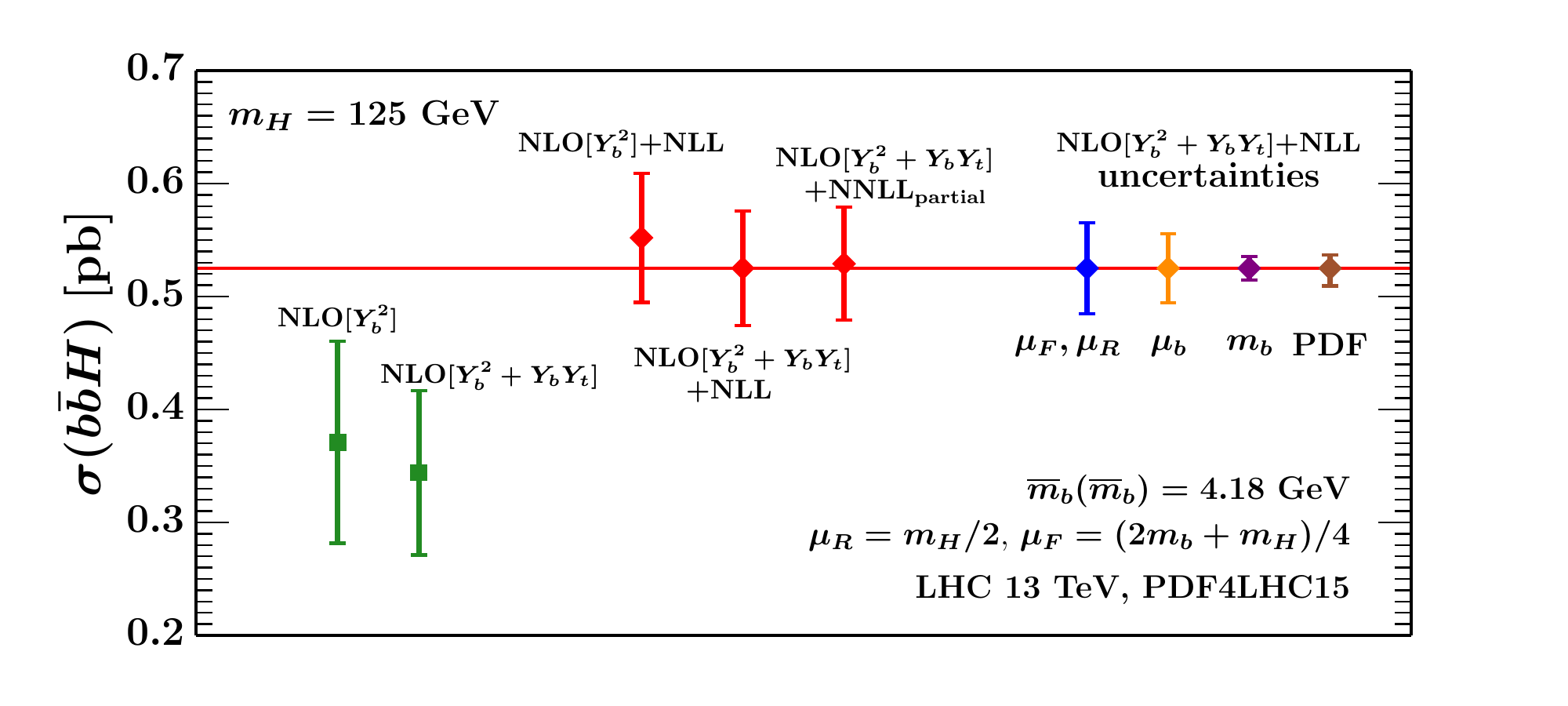}
\caption{Comparison of the cross sections for $m_H=125$~GeV at the 13~TeV LHC at \NLO, \NLOt\ without resummation and at \NLO+NLL, \NLOt+NLL, and \NLOt+NNLL$_\pa$. A full breakdown of the uncertainties at \NLOt+NLL is also shown.}
\label{fig:mh125}
\end{figure}

\begin{figure}[t]
\centering
\includegraphics[scale=0.75]{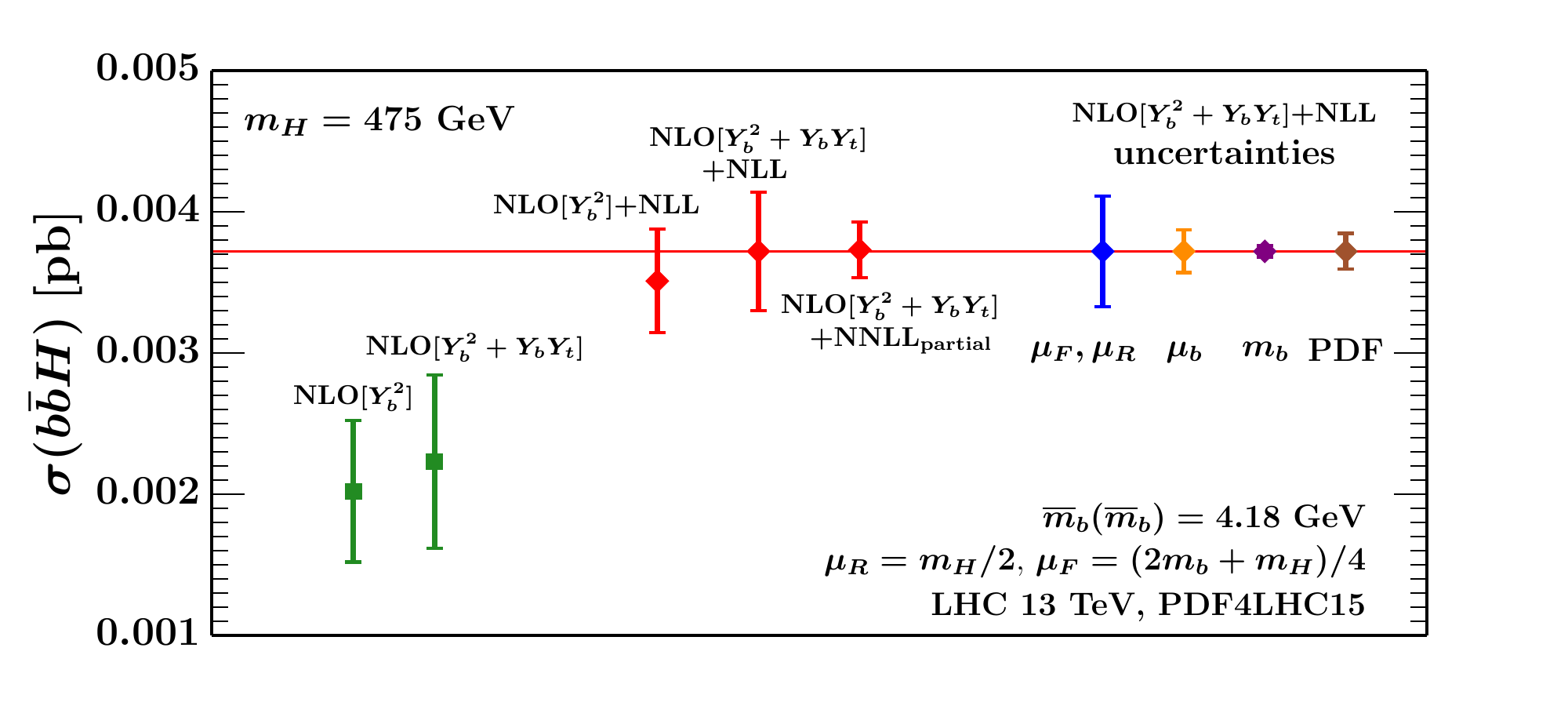}
\caption{Comparison of the cross sections for $m_H=475$~GeV at the 13~TeV LHC at \NLO, \NLOt\ without resummation and at \NLO+NLL, \NLOt+NLL, and \NLOt+NNLL$_\pa$. A full breakdown of the uncertainties at \NLOt+NLL is also shown.}
\label{fig:mh475}
\end{figure}

\begin{figure}[t]
\centering
\includegraphics[scale=0.75]{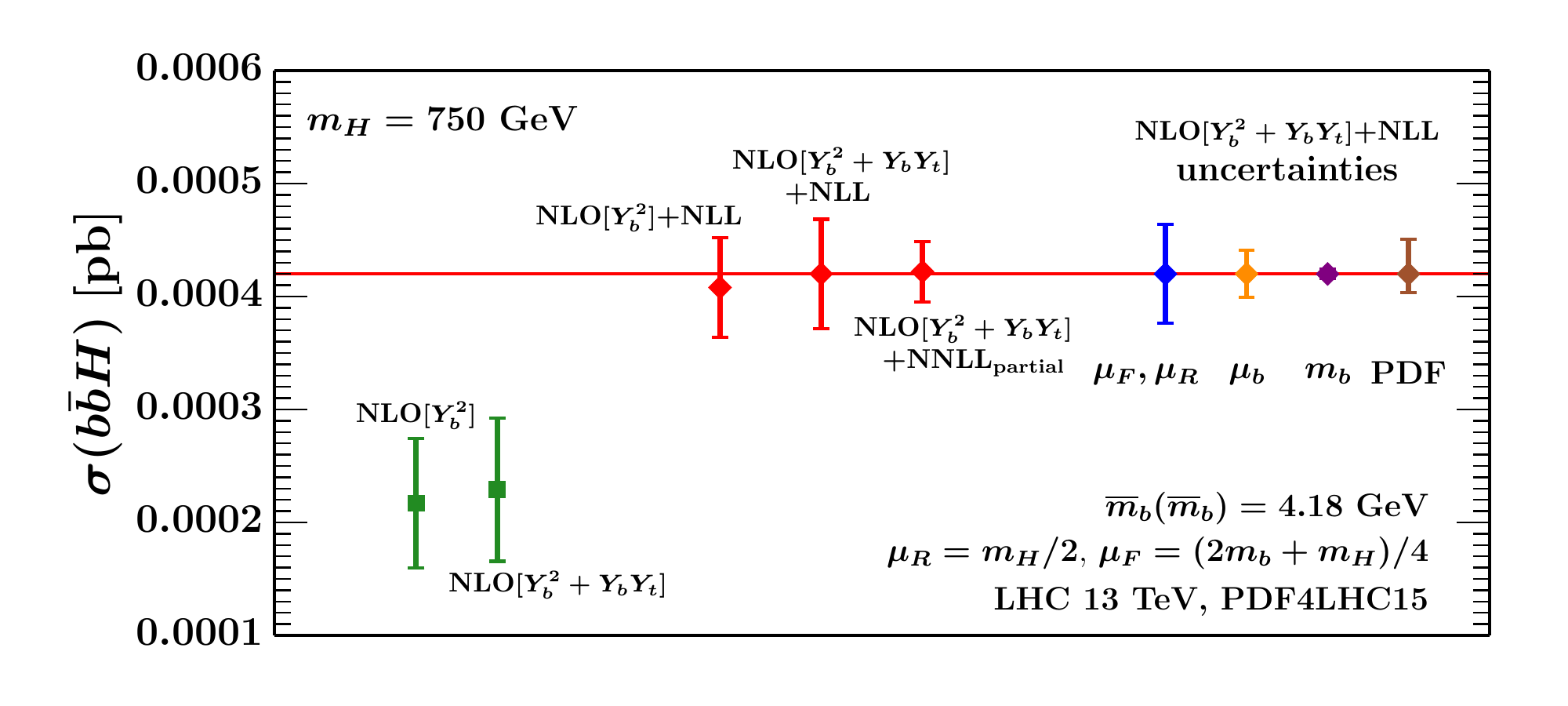}
\caption{Comparison of the cross sections for $m_H=750$~GeV at the 13~TeV LHC at \NLO, \NLOt\ without resummation and at \NLO+NLL, \NLOt+NLL, and \NLOt+NNLL$_\pa$. A full breakdown of the uncertainties at \NLOt+NLL is also shown.}
\label{fig:mh750}
\end{figure}

In this section, we present our numerical results for the inclusive $b\bar{b}H$ cross section
for values of the Higgs boson mass in the range $m_H \in [50,750]\GeV$.
We also consider values of the Higgs mass different from $m_H=125\GeV$, as these
are relevant for BSM scenarios in which the Higgs coupling to the bottom quark is enhanced.
For simplicity we always use SM couplings --
the cross section in many BSM scenarios can be obtained by rescaling $Y_b$ by an appropriate 
model-dependent factor~\cite{Carena:1999py,Guasch:2003cv,Dawson:2005vi}.

We always use the simplified implementation
without the strict expansion, which should still be valid at the lowest considered value of $m_H$. The highest value 
is chosen only semirandomly~\cite{ATLAS-CONF-2016-018, CMS:2016owr}.
The results for the cross section at \NLO+NLL, \NLOt+NLL, \NLO+NNLL$_\pa$,
and \NLOt+NNLL$_\pa$ are given in tables~\ref{tab:nlonll}, \ref{tab:nlonllybyt}, \ref{tab:nlonnllp}, and
\ref{tab:nlonnllpybyt}, respectively.
The precise definitions of the different orders are discussed in \sec{nlonll}.
In the cross section tables we give the central value for the cross section together with the full
breakdown of the perturbative uncertainties due to $\{ \mu_F,\mu_R \}$ and $\mu_b$ as well as the
parametric uncertainties due to $m_b$ and PDFs.

For each of the fixed-order ($\muf$ and $\mur$), resummation ($\mub$),
and parametric $m_b$ uncertainties we use the absolute value of the maximum deviation from the central result as the symmetric
uncertainty. The PDF uncertainty is computed according to the PDF4LHC15 prescription as described in \subsec{pdf} and is kept asymmetric.
As discussed in \subsec{scales}, the full perturbative uncertainty is obtained by adding the
fixed-order and resummation uncertainties in quadrature.
If one is only interested in the total $b\bar bH$ cross section, the perturbative and parametric uncertainties
can be added in quadrature. In a more complicated setup, e.g., a global fit, the total perturbative, $m_b$, and PDF uncertainties
should be treated as independent uncertainty sources (e.g.\ they correspond to independent nuisance parameters). This allows one to properly take into account their correlations with other predictions affected by the same physical uncertainty sources. For example, the parametric uncertainty due to $m_b$ or $Y_b$ should be correlated with the corresponding parametric uncertainty when bottom loop effects are included in $ggH$.

Our results are also illustrated in three figures.
In \fig{mhscan}, we show the LO+LL,\footnote{The LO+LL result here is consistently computed with NLO PDFs.}
\NLOt+NLL, and \NLOt+NNLL$_\pa$ cross sections as a function
of the Higgs mass. Here, the uncertainty bands show the total perturbative uncertainty adding in quadrature
the $\{ \mu_F,\mu_R \}$ and $\mu_b$ uncertainties.
The important message that follows from this figure is that we can see an excellent perturbative convergence between the orders.
Additionally, the band for the \NLOt+NNLL$_\pa$ cross section is fully included within the \NLOt+NLL band, and the central values
for the two results are almost identical, which is a nice feature of our central scale choices.
This pattern also gives us a good degree of confidence in the method we use to estimate the perturbative uncertainties.

These conclusions are unchanged when the $Y_bY_t$ interference terms are omitted.
In the lower panel of the figure we show the ratio of the central \NLOt+NLL over the central \NLO+NLL to illustrate the
numerical effect of the $Y_bY_t$ interference terms on the matched result.
We see that the effect of adding the interference term is moderate but clearly noticeable in the numerical results,
giving a negative contribution for $m_H\lesssim 300$~GeV, and a positive one for larger masses.
(The small fluctuations are due to the finite integration statistics in MC@NLO when including the interference terms.)

Note that for large Higgs masses the uncertainties at NLO+NNLL$_\pa$ get significantly smaller
than at NLO+NLL. As discussed in \subsec{nlonnllp}, this is expected since for large $m_H$
the 5FS perturbative counting is more appropriate and in this limit the NLO+NNLL$_\pa$ formally improves the accuracy of
the result, becoming as accurate as the full NNLO 5FS cross section.
For smaller values around the physical Higgs mass of $m_H = 125\GeV$, our counting is more appropriate, and only the full NNLO+NNLL
should be regarded as a complete next-order result. Therefore, in this region the larger NLO+NLL uncertainty should be regarded as a safer
estimate of the residual theory uncertainty.
This is also nicely confirmed by the fact that in this range the NLO+NNLL$_\pa$ uncertainty is essentially as large as at NLO+NLL.

In figures~\ref{fig:mh125}, \ref{fig:mh475}, and \ref{fig:mh750} we give a visual breakdown of the results for $m_H=125$~GeV,
$m_H=475$~GeV, and $m_H = 750\GeV$. The matched results and uncertainty contributions are equivalent to the numbers provided in the tables.
In addition, we also give a comparison with the corresponding pure fixed-order
results at NLO with and without $Y_bY_t$ terms (green points).%
\footnote{These are the fixed \NLO\ and \NLOt\ results that are contained in our NLO+NLL result and that would be obtained
if we were to take $\mu_b \to \mu_F$ to turn off the resummation. These are still consistently
computed with $n_f = 5$ running for gluon and light-quark PDFs and $\as$ and are thus not numerically identical to the usual 4FS result
that uses $n_f = 4$ running everywhere. The difference is however of higher order in the 4FS expansion.}
By comparing the green NLO with the red NLO+NLL points we can see that the effects of the resummation are significant,
resulting in a $\sim30\%$ increase for $m_H=125$~GeV, and even more at the higher Higgs masses,
and moreoever this effect is not covered by the fixed-order scale variation band. This clearly shows that the resummation of $b$-quark collinear logarithms cannot be neglected for these mass values.
Comparing the red points for \NLOt+NLL and \NLOt+NNLL$_\pa$ we see the same features as we saw
in fig.~\ref{fig:mhscan}.

The four points on the right of figures~\ref{fig:mh125}, \ref{fig:mh475}, and \ref{fig:mh750}
show the breakdown of the uncertainties for our default result at \NLOt+NLL.
Clearly, the fixed-order uncertainty estimated by the $\muf$ and $\mur$ variations
is the largest source of uncertainty in the predictions, both for moderate as well as large
values of $m_H$. The orange points show the resummation uncertainty estimated from the $\mu_b$ variation, which,
although smaller than the fixed-order uncertainty by roughly a factor of two, is nevertheless not negligible.
We also recall that at the lower LO+LL order the resummation uncertainty plays a significant role,
as was shown in \mycite{Bonvini:2015pxa}.

The parametric uncertainties due to $m_b$ (purple points) and PDFs (brown points) are subdominant.
The $m_b$ uncertainties are small around $2\%$ and decrease for higher Higgs masses to around $1\%$.
The (asymmetric) PDF uncertainties, computed as described in \subsec{pdf},
are around $4\%$ and smaller than both the $\{\muf, \mur\}$ and $\mu_b$ uncertainties.

One point to emphasize is that the uncertainty which is solely due to the parametric uncertainty in $m_b$ is
much smaller than the perturbative $\mu_b$ uncertainty, especially for larger Higgs masses.
This shows the importance of distinguishing these two effects. If we were to identify $\mub\equiv m_b$,
as is commonly done, we would be left to choose between two undesirable options.
Either we could vary their common value in the range \eq{mbpole-variation}, which would essentially set the $\mub$ uncertainty to zero. Or, we could vary their common value in a much larger
range to account for the $\mub$ uncertainty, which however would be unjustified for $m_b$ and blow up the parametric $m_b$ uncertainty.
In contrast, by identifying and separating these two uncertainty sources, we are able to properly estimate each of them.

\section{Conclusions}
\label{sec:conclusions}

We have presented state-of-the-art predictions for the $b\bar bH$ cross section at 13~TeV obtained
from a matched calculation~\cite{Bonvini:2015pxa} that consistently combines the fixed-order (4FS) contributions
(which include the full $b$-quark mass dependence)
with the all-order resummation of collinear logarithms of $m_b/m_H$.
We provide results with and without including the effect of the interference of top-loop induced Higgs production process
with the pure bottom-induced production proportional to $Y_bY_t$.
We also study the effect of two-loop contributions that formally contribute at NNLL order,
finding that they are small and their effect fully captured by the uncertainty of our default NLO+NLL result.

We perform a detailed study of several sources of uncertainty in our results, both theoretical and parametric.
The perturbative uncertainty from missing higher orders is estimated by varying the hard scales $\muf$ and $\mur$,
as well as the resummation scale $\mub$, which represents the threshold scale at which 4FS evolution is matched
to 5FS evolution in the PDFs.
We consider our resulting theory uncertainty as a reliable estimate, which is neither aggressive nor overly conservative.
Furthermore, the parametric uncertainties due to the $b$-quark mass value and PDFs are evaluated.
In particular, we discuss how to disentangle the unphysical dependence on
the $b$-quark matching scale $\mub$ from the purely parametric dependence on the $b$-quark mass $m_b$,
which requires the construction of dedicated 5F PDFs.

Our methodology to compute the matched prediction and to evaluate its uncertainties
can be readily applied to other heavy-quark-initiated processes at the LHC.
The code for our matched predictions will be available at \href{http://www.ge.infn.it/~bonvini/bbh}{\texttt{http://www.ge.infn.it/$\sim$bonvini/bbh}}.
Our results represent the currently most complete predictions for the $b\bar b H$ cross section
in the Standard Model and we are looking forward to a first measurement of this process during the coming LHC Run 2.

\begin{acknowledgments}
We thank Stefano Forte, Robert Harlander, Stefan Liebler, Davide Napoletano, Michael Spira, Robert Thorne, 
Maria Ubiali, and Marius Wiesemann for useful discussions.
The work of MB is supported by an European Research Council Starting Grant ``PDF4BSM: Parton Distributions in the Higgs Boson Era''.
The work of AP is supported by the UK Science and Technology Facilities Council [grant ST/L002760/1].  
The work of FT was supported by the DFG Emmy-Noether Grant No.\ TA 867/1-1.
\end{acknowledgments}

\phantomsection
\addcontentsline{toc}{section}{References}

\bibliographystyle{../jhep}
\bibliography{bbh_refs}

\end{document}